\documentclass[useAMS,usenatbib]{mn2e}
\usepackage{setspace}
\usepackage{graphicx}
\usepackage{caption}
\usepackage{subcaption}
\usepackage{lscape}

\begin{document}
\title[ Investigation of subgiants and giants from ASAS]{Orbital and physical parameters of eclipsing binaries from the ASAS catalogue - V. Investigation of subgiants and giants: the case of ASAS J010538-8003.7, ASAS J182510-2435.5, and V1980 Sgr\thanks{Based on observations made with ESO telescopes at the La Silla-Paranal Observatory under programme ID: 087.C-0012 and through  CNTAC proposals CN-2011B-21 and CN-2012A-21}\thanks{Based in part on data collected at Subaru Telescope, which is operated by the National Astronomical Observatory of Japan, via the time exchange program between Subaru and the Gemini Observatory}}
\author[M. Ratajczak et al.] {M. Ratajczak$^{1}$\thanks{E-mail:
milena@ncac.torun.pl}, K. G. He{\l}miniak$^{2,1}$, M. Konacki$^{1,3}$,  and A. Jord\'an$^{2}$\\
$^{1}$Nicolaus Copernicus Astronomical Center, Department of Astrophysics, ul. Rabia\'{n}ska 8, 87-100 Toru\'{n}, Poland\\
$^{2}$Departamento de Astronom\'{i}a y Astrof\'{i}sica,
 Vicu\~{n}a Mackenna 4860, 782-0436 Macul, Santiago, Chile\\
$^{3}$Astronomical Observatory, A. Mickiewicz University, ul. S{\l}oneczna 36, 60-286 Pozna\'{n}, Poland}

\date{Accepted... Received...}

\pagerange{\pageref{firstpage}--\pageref{lastpage}} \pubyear{2013}

\maketitle

\label{firstpage}

\begin{abstract}

We present absolute physical and orbital parameters for three double-lined detached eclipsing binary systems from the \textit{All Sky Automated Survey} (ASAS) catalogue with subgiant and giant components. These
parameters were derived from archival \textit{V}-band ASAS photometry and new radial velocities. The radial velocities were calculated from high quality optical spectra we obtained with the 8.2~m Subaru/HDS, ESO 3.6~m/HARPS, 1.9~m Radcliffe/GIRAFFE, CTIO 1.5~m/CHIRON,  and 1.2~m Euler/CORALIE using the two-dimensional cross-correlation technique (TODCOR) and synthetic template spectra chosen for every system separately as references.

The physical and orbital parameters of the systems were derived with the \textsc{phoebe} and \textsc{jktebop} codes. We checked the evolutionary status of the systems with several sets of isochrones and determined distances for each system. The derived uncertainties for individual masses of ASAS J010538-8003.7, ASAS J182510-2435.5, and V1980 Sgr components vary from 0.7$\%$ to 3.6$\%$ while the radii are in the range 1$\%$ to 24$\%$ . For all of the investigated systems such a detailed orbital and physical analysis is presented for the first time.
 
\end{abstract}


\section{Introduction}

Stellar astrophysics, in many aspects, is based on precisely determined fundamental parameters of stars such as their masses and radii. With a few exceptions, only the eclipsing binaries of SB2 type enable us to determine directly those parameters with the required precision. Hence such objects allow us to test the predictions of theoretical stellar evolution models.  Recent review papers about detached eclipsing binaries \citep[e.g.,][]{tor10} present the comparison between observations and theoretical models for late spectral type objects. Systems containing stars in advanced stages of evolution (e.g., red giants) are much less investigated than, for example, low-mass binaries (which consist of dwarfs), also considered to be poorly studied \citep[e.g.,][]{rib08, hel11a}. 

So far, only a dozen of stars with subgiant and giant components in eclipsing binaries have been characterized in the Milky Way with the required masses and radii accuracy of 3\% \citep[upper limit by][]{tor10}, which is indicative of their usefulness as test beds for theoretical models (e.g. TZ For, MY Cyg, HY Vir). Worth mentioning is also very informative system AI Phe  \citep{and88, hel09} --  a rare example of differential evolution. The combination of a main sequence star with one that is already at the lower giant branch is ideal for an empirical verification of the stellar evolution theoretical models. Additionally, there are 12 extragalactic systems with well characterized giants discovered in LMC \citep{pie09,pie10,pie11,pie13} and SMC \citep{gra12}. The analysis of these systems shows that our knowledge of the more advanced stages of stellar evolution (following the main sequence phase) is incomplete.

An increasing number of observations of late type binaries have intensified theoretical work on stellar structure and evolution. However, we are still lacking high quality data. Using high resolution spectrographs and available photometric measurements we can increase the sparse sample of giants with accurately determined parameters. 

In this paper we present the latest results of our on-going spectroscopic survey \citep{hel09,hel11a,hel11b,hel12} of eclipsing binaries from the \textit{All Sky Automated Survey}\footnote{\texttt{http://www.astrouw.edu.pl/asas/?page=acvs}} \citep[hereafter ASAS,][]{poj02, pac06}. We focus on three systems where both components are subgiant or giant stars. First we describe our targets, then its data collection and analysis, and finally the results we obtained. Section 6 contains the discussion about the evolutionary status of the systems, and age and distance determination, while Section 7 summarizes the main conclusions.

\section{Targets}

The observing strategy included the selection of detached eclipsing binaries from the extensive \textit{ASAS Catalogue of Variable Stars} \citep[ACVS;][]{poj02} and a spectroscopic campaign to infer the evolutionary status of every component and determine their physical and orbital parameters. The systems were selected on the basis of the following criteria: P $>$ 8 days, change in brightness $<$ 1~mag, \textit{V}-\textit{K} $>$ 1~mag, in order to search for detached, non-Algol systems with components of solar radius or larger. The analyzed sample include the binary systems: ASAS J010538-8003.7, ASAS J182510-2435.5, and V1980 Sgr. 

\subsection{ASAS J010538-8003.7}
The eclipsing binary ASAS J010538-8003.7 (CD-80 28, 2MASS J01053817-8003409, TYC 9355-177-1, hereafter ASAS-010538) is classified as an eclipsing detached binary system (DEB) in the ACVS. Its apparent \textit{V} magnitude is 10.1 \citep{poj02}, and the amplitude of photometric variations in \textit{V}-band is 0.44 mag. The system was briefly analyzed by \cite{hel09} who pointed out that the binary components are subgiants and suggested its further investigation.

\subsection{ASAS J182510-2435.5}
The eclipsing binary ASAS J182510-2435.5 (TYC 6861-523-1, hereafter ASAS-182510) is classified as a DEB in the ACVS. Its apparent \textit{V} magnitude is 10.87 \citep{hog00}. However, in the ACVS it is estimated to be 10.56. The reason for that significant difference is the additional visual component that contributes to the total brightness of the system (third light) inside an aperture used by the ASAS pipeline to measure the stellar flux. The amplitude of photometric variations in the ASAS \textit{V}-band curve is 0.3 mag. There is no analysis of this system in the literature so far.

\subsection{V1980 Sgr}
The eclipsing binary V1980 Sgr, known also as ASAS J182525-2510.7 (CD-25 13101, HD 315626, TYC 6861-2115-1, hereafter ASAS-182525) was discovered in 1962 \citep{hof62} and previously considered to be a semi-detached system. However, in the ACVS it is classified as a detached system. Its apparent \textit{V} magnitude is 10.2 \citep{poj02}, and the amplitude of photometric variations in \textit{V}-band is 0.57 mag. Additionally, significant, time-varying, out-of-eclipse modulation is clearly visible. The system has not been spectrosopically investigated before.

\section{Observations}

\subsection{Spectroscopy}

All of the selected systems are double-lined spectroscopic binaries (SB2). In order to measure radial velocities (RVs) of both components of all of the systems we carried out observations with a wide range of spectrographs. We used the following telescopes and instruments: the 8.2~m Subaru telescope and the High Dispersion Spectrograph (HDS; R$\sim$60\,000), the 3.6~m ESO telescope equipped with the High Accuracy Radial velocity Planet Searcher (HARPS) spectrograph \citep[][R$\sim$115\,000]{may03}, the 1.9~m Radcliffe telescope and the GIRAFFE spectrograph (R$\sim$40\,000), the 1.5~m CTIO telescope equipped with the CHIRON\footnote{Operated by the SMARTS Consortium} spectrograph \citep[][Tokovinin et al. 2013, in preparation, R$\sim$90\,000]{sch12}, and the 1.2~m Euler telescope with the CORALIE spectrograph \citep[][R$\sim$60\,000]{que01}.

\begin{figure}
\includegraphics[scale=0.35,angle=270]{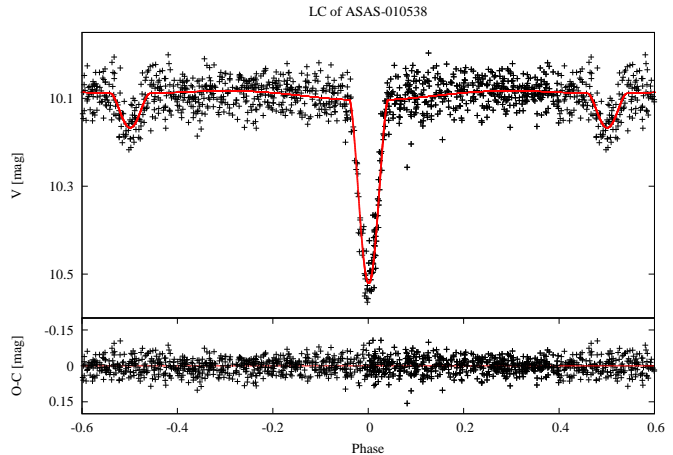}
\caption{The observed $V$ light curve of ASAS-010538 from ACVS catalogue phased with the period P=8.069 d and the best-fitting model. The residuals are shown in the lower panel.
}\label{fig_lc}
\end{figure}

\begin{figure}
\includegraphics[scale=0.35,angle=270]{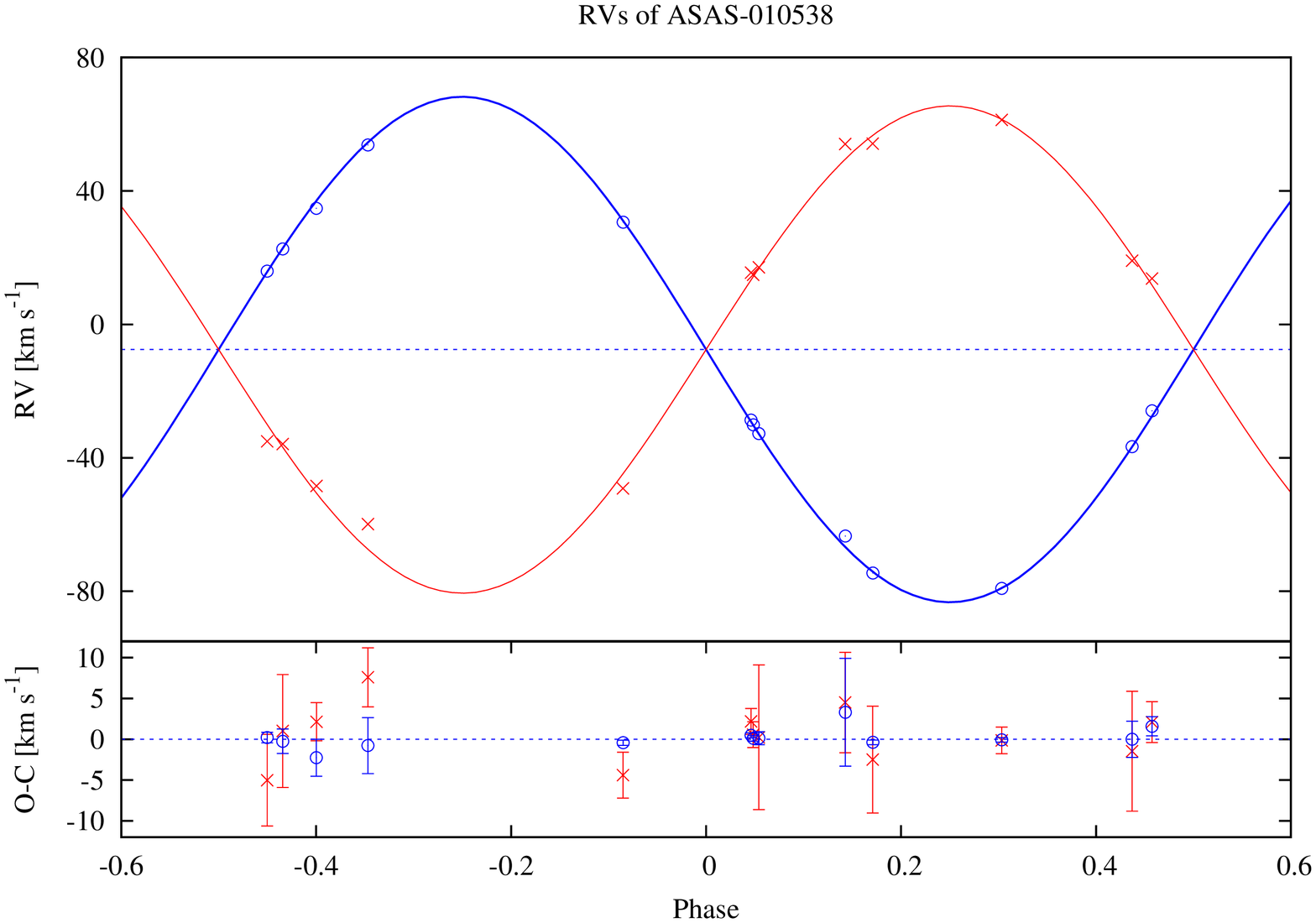}
\caption{The observed RVs of both components of ASAS-010538 with their best-fitting Keperian models and O-Cs with corresponding \textit{rms} (lower panel). Circles represent measurements of the primary and crosses measurements of the secondary.} 
\label{fig_rv}
\end{figure}

\begin{figure}
\includegraphics[scale=0.35,angle=270]{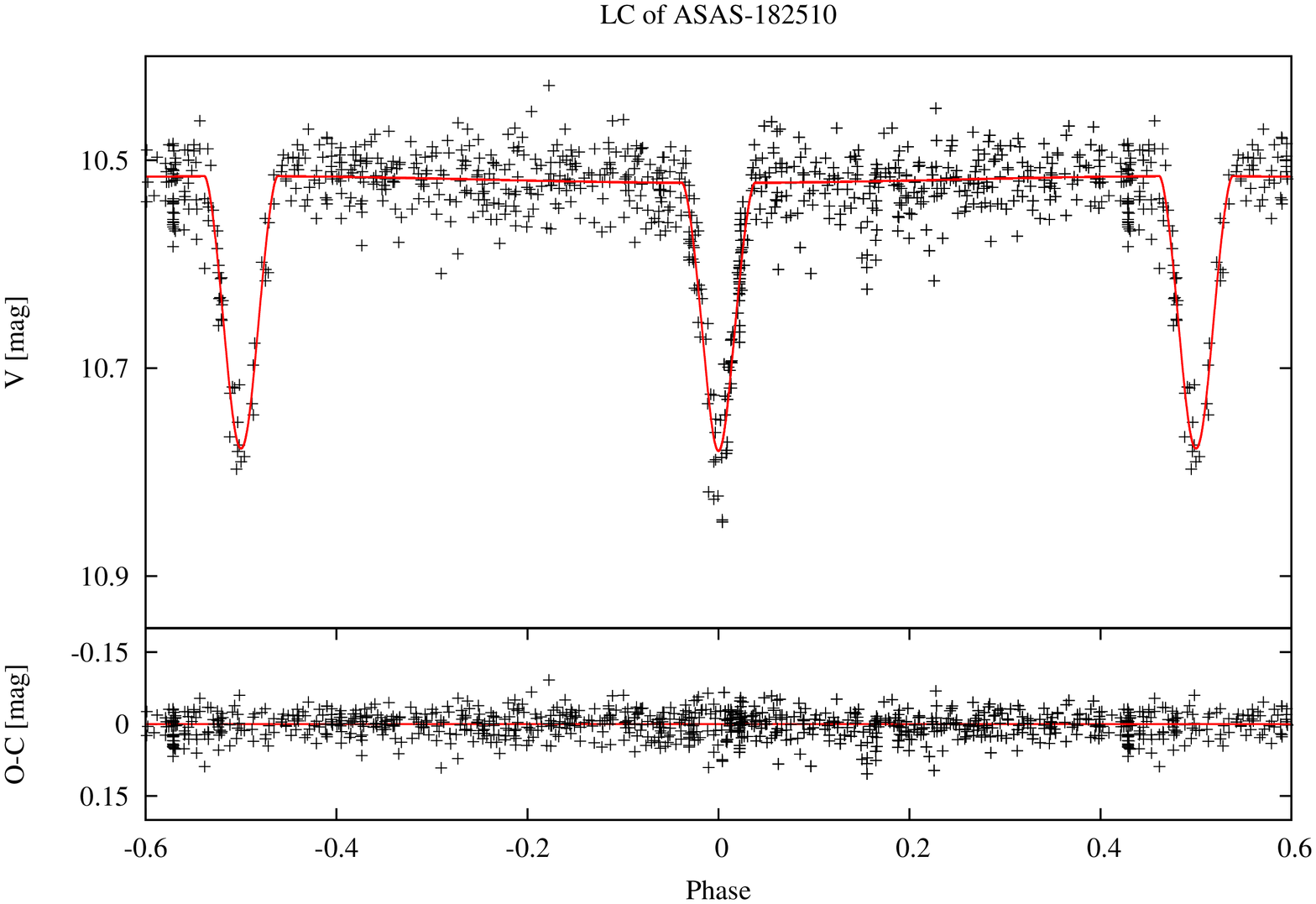}
\caption{The observed $V$ light curve of ASAS-182510 from ACVS catalogue phased with the period P=86.648 d and the best-fitting model. The residuals are shown in the lower panel.
}\label{fig_lc_182510}
\end{figure}

\begin{figure}
\includegraphics[scale=0.35,angle=270]{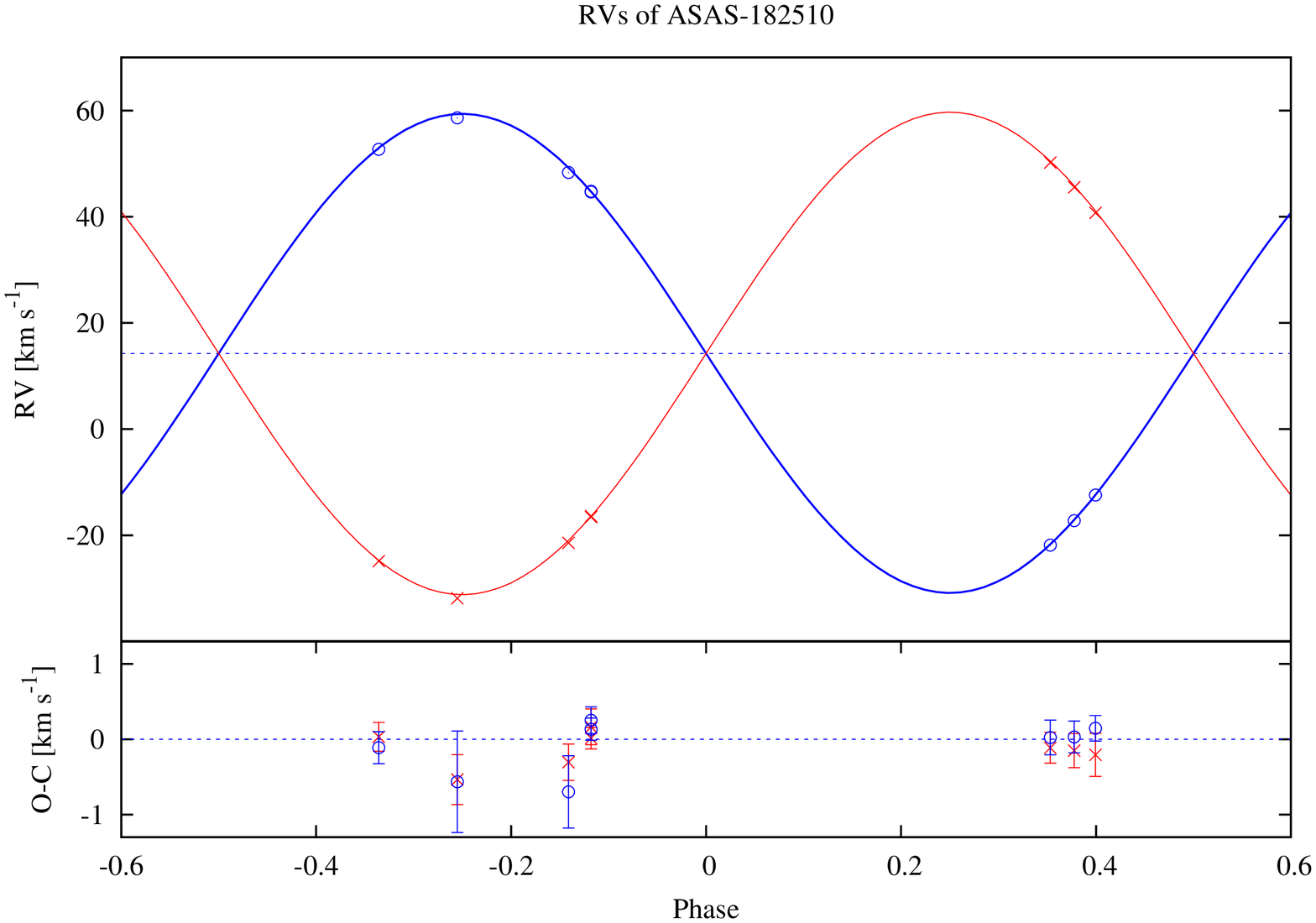}
\caption{The observed RVs of both components of ASAS-182510 with their best-fitting Keperian models and O-Cs with corresponding \textit{rms} (lower panel). Circles represent measurements of the primary and crosses measurements of the secondary.} 
\label{fig_rv_182510}
\end{figure}

\begin{figure}
\includegraphics[scale=0.35,angle=270]{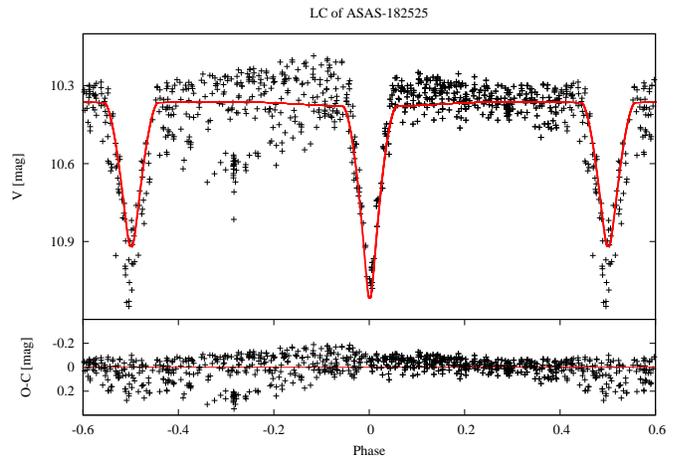}
\caption{The whole observed $V$ light curve of ASAS-182525 from ACVS catalogue phased with the period P=40.506 d and the best-fitting model for the base season (see text). The residuals are shown in the lower panel.
}\label{fig_lc_182525}
\end{figure}

\begin{figure}
\includegraphics[scale=0.35,angle=270]{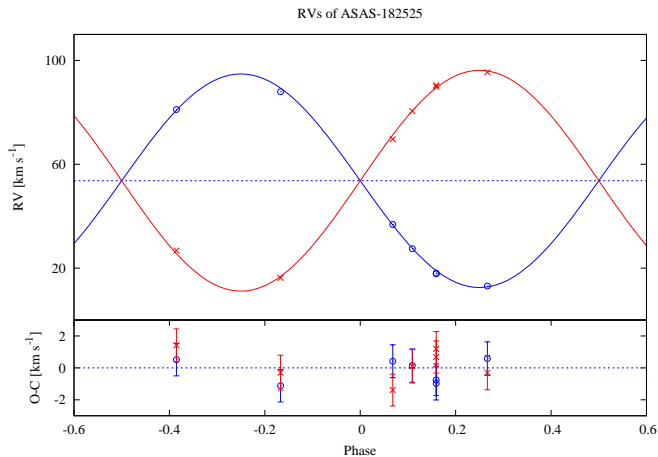}
\caption{The observed RVs of both components of ASAS-182525 with their best-fitting Keperian models and O-Cs with corresponding \textit{rms} (lower panel). Circles represent measurements of the primary and crosses measurements of the secondary.} 
\label{fig_rv_182525}
\end{figure}

\subsubsection{ASAS-010538}
To obtain a new spectroscopic solution for ASAS-010538 system we used 13 RV measurements for both components. Four of them were already published \citep{hel09}, but we rejected one measurement due to low SNR and high RV errors. Three additional RV measurements were derived from GIRAFFE spectra collected during an observing run in September 2010, five more were obtained using the HARPS spectrograph in June and September, 2011, two more come from spectra collected in August 2011 with CHIRON in service mode.  

\subsubsection{ASAS-182510}
We obtained 8 RV measurements for both components of ASAS-182510 system. Four of them were derived from CHIRON spectra collected  in July and September, 2011 (service mode), two were obtained using CORALIE spectrograph in September 2011 and May 2012, and two RV measurements were derived from two CCD chips (blue and red) on which the spectrum taken by the HDS spectrograph at the Subaru telescope (August 2011) was recorded. 

\subsubsection{ASAS-182525}
We obtained 7 RV measurements for both components of ASAS-182525 system. Two of them were derived from CHIRON spectra collected  in July 2011 (service mode), three were obtained using the CORALIE spectrograph in September 2011 and May 2012, and two RV measurements come from the two separate CCD chips of the HDS spectrograph (August 2011). 

\subsection{Photometry}

We used the\textit{ V}-band measurements of the selected system  from the ACVS. The data phase coverage is complete for all of the systems, and the data span is around 9 years (HJD$\sim$2451900-2455200). We used 1089, 797, and 769 measurements for ASAS-010538, ASAS-182510, and ASAS-182525, respectively. For ASAS-010538 this is more than used in the solution presented by \cite{hel09}.

\section{Analysis}

\subsection{Radial Velocities}
For CCD reduction, spectrum extraction and wavelength calibration for data from GIRAFFE and HDS spectrographs we used standard \textsc{iraf}\footnote{\textsc{iraf} is written and supported by the \textsc{iraf} programming group at the National Optical Astronomy Observatories (NOAO) in Tucson, AZ. NOAO is operated by the Association of Universities for Research in  Astronomy (AURA), Inc. under cooperative agreement with the  National Science Foundation. \texttt{http://iraf.noao.edu/}} procedures for echelle spectra. The wavelength calibration was done in the standard manner with Thorium-Argon lamp frames taken before and after object exposures (GIRAFFE) or at the beginning and at the end of observing nights (HDS). We used the \textsc{iraf} \textit{rvsao.bcvcorr} task for barycentric velocity corrections.

The basic reduction, spectral extraction, and wavelength calibration of the spectra from the CHIRON spectrograph were done with the pipeline developed at Yale University (Tokovinin et al. 2012, in preparation). The wavelength calibration was based on the Thorium-Argon lamp exposure taken just before the science exposure with the same spectrograph settings. As the pipeline does not include barycentric velocity corrections, we used the \textit{rvsao.bcvcorr} task for that purpose.  

Data from the HARPS spectrograph were reduced using dedicated software (\textit{Data Reduction Software}, \textit{DRS}) which allows spectra to be reduced in near real-time. The pipeline performs barycentric corrections. The instrument works in a simultaneous wavelength calibration mode, so Thorium-Argon exposures are taken simultaneously with the exposure of an object. 

Spectra from the CORALIE  spectrograph were also taken in a simultaneous Thorium-Argon mode, where one of the fibres observes a target and the other an arc lamp. For the data reduction, calibration, and barycentric corrections we used an automated pipeline developed at Pontificia Universidad Cat\'{o}lica de Chile. It it briefly described in \cite{pen13}, a full description will be presented in an upcoming work (Jordan et al. 2013, in prep.).

The RVs were calculated using our own implementation of the two dimensional cross-correlation technique \citep[TODCOR;][]{zuc94} using as references various synthetic spectra computed with the \textsc{atlas9} and \textsc{atlas12} codes \citep{kur92}. The formal RV errors were computed from the bootstrap analysis of TODCOR correlation maps created by adding selected single-order maps. To obtain a best-fit with reduced $\chi^{2}$ $\approx$ 1 to our RV data, we multiplied the formal errors by an appropriate factor.

The RV measurements, their final errors (formal errors multiplied by a calculated factor), and \textit{O}-\textit{C}s are collected in Tables \ref{RV_010538}--\ref{RV_182525} in Appendix A. The tables also include the exposure times and a signal-to-noise ratio (SNR) per collapsed pixel at $\lambda$=5\,500 \AA ~of each spectrum.

As we work with heterogenous datasets, we checked if there is no differences between RV zero points from spectrographs we use. To deal with that, we fit an additional parameter, which allows to compensate for different zero points in the procedure we use to obtain preliminary results from RV measurements. Initially, we set the parameter free and concluded the difference in RV zero points between spectrographs is insignificant and has no influence on the final results, thus the final fits were done with that parameter fixed to 0.

\subsection{Modelling}

Our RV measurements were combined with the ASAS photometry to derive absolute orbital and physical parameters of the systems. To accomplish these goals we proceeded as follows:
\begin{itemize}
\item To obtain a preliminary solution from RV measurements we used a procedure that fits a double-Keplerian RV orbit and minimizes the $\chi^{2}$ function with a Levenberg-Marquardt algorithm. Every system components' mass ratio was derived in this manner and applied in the further analysis. \\
\item The light curve modelling was done with \textsc{jktebop} \citep{sou04a, sou04b} based on \textsc{ebop} \citep[\textit{Eclipsing Binary Orbit Program};][]{pop81, etz81}. The code fits a geometric model of a detached eclipsing binary to a light curve in order to derive components' and systems' parameters. \textsc{jktebop} includes also an extensive Monte Carlo error analysis algorithm which was used to determine robust uncertainties in the parameters which the program derives. We checked the eccentricity for every system and concluded that all of the orbits are consistent with being circular, so we kept the eccentricity fixed at 0 for the further analysis. The parameters like the period and time of minimum ($P$ and $T_{0}$ were much more precise than the quantities from the ASAS analysis presented in ACVS), the inclination were calculated and applied to the repeated RV - Keplerian orbit fitting procedure. We used \textsc{jktebop} to derive the fractional stellar radii of the components. \\
\item To estimate the semimajor axis, systemic velocity, luminosities, gravitational potentials, and effective temperatures of system components and deal with spots (in the case of ASAS-182525) we used the \textsc{phoebe} code \citep[\textit{Physics Of Eclipsing Binaries};][]{prs05} - an implementation of the Wilson-Devinney code \citep{wil71} which uses the computed gravitational potential of each star to determine the surface gravity and effective temperatures.\\
\item To derive absolute values and their uncertainties we used the procedure \textsc{jktabsdim} \citep{sou04a, sou04b}. It calculates the absolute dimensions, related quantities, and the distance of the detached eclipsing binary from the results of a radial velocity and light curve analysis.\\
\end{itemize}

\begin{table*}
\caption{Orbital and physical parameters of ASAS-010538, ASAS-182510, and ASAS-182525}
\label{tab_orb}
\begin{tabular}{lrlrlrl}
\hline
\hline
Parameter & \multicolumn{2}{c}{ASAS-010538} & \multicolumn{2}{c}{ASAS-182510} & \multicolumn{2}{c}{ASAS-182525}\\
\hline
$P$ [d] & 8.069388 & $\pm$0.000021 &  86.6486 & $\pm$0.0047 &  40.50556 & $\pm$0.00055 \\
$T_{0}$ [JD-2450000]	& 1873.4519 & $\pm$0.0051 & 2063.304  &$\pm$0.073 & 2005.297 &$\pm$0.021\\
$K_{1}$ [km~s$^{-1}$] 	& 75.74 & $\pm$0.26 & 45.45 &  $\pm$0.12 & 42.53 & $\pm$0.59\\
$K_{2}$ [km~s$^{-1}$] 	& 73.0 & $\pm$1.3 & 45.12 & $\pm$0.13 & 41.14 & $\pm$ 0.57\\
$\gamma$ [km~s$^{-1}$]& -7.597 &  $\pm$0.053 & 14.237 & $\pm$0.014 & 53.541 & $\pm$0.048 \\
$q$ 		& 1.04 &  $\pm$0.02 & 1.007 & $\pm$0.004 & 1.03 & $\pm$0.02 \\
$e$ 		 	&  0.0  & (fixed) & 0.0 & (fixed) & 0.0 & (fixed) \\
$i$ & 79.97 &  $\pm$0.65 & 85.61 & $\pm$0.86 & 84.2 & $\pm$1.4\\
$a$ [R$_\odot$] & 24.201 & $\pm$0.028 & 152.663 & $\pm$0.066 &  67.29 & $\pm$0.11\\
$M_{1}$ [M$_\odot$] & 1.415 &  $\pm$0.051 & 3.353 & $\pm$0.025 & 1.227 & $\pm$0.039\\
$M_{2}$ [M$_\odot$] & 1.467 & $\pm$0.029 & 3.377 & $\pm$0.024 & 1.269 & $\pm$0.040\\
$R_{1}$ [R$_\odot$] & 2.02 & $\pm$0.29 & 23.44 & $\pm$2.97 & 12.97 & $\pm$0.97\\
$R_{2}$ [R$_\odot$] & 5.27 & $\pm$0.16 & 15.26 & $\pm$3.89 & 12.74 & $\pm$1.06\\
$v_{rot1}$ [km~s$^{-1}$] & 12.6 & $\pm$1.9 & 13.7 & $\pm$1.7 & 16.2 & $\pm$1.2\\
$v_{rot2}$ [km~s$^{-1}$] & 33.1 & $\pm$1.0 & 8.9 & $\pm$2.3 & 15.9 & $\pm$1.3\\
$T_{1}$ [K] & 6156 & $\pm$176 & 4830 & $\pm$97 & 4783 & $\pm$82\\
$T_{2}$ [K] & 4889  & $\pm$98 & 4800 & $\pm$107 & 4600 & $\pm$163\\
\hline
\end{tabular}
\end{table*}

\section{Results}
The results of our modelling of the selected systems are shown in Table \ref{tab_orb}. The physical and orbital parameters are given with their 1~$\sigma$ uncertainties. The primary component is defined as a star eclipsed during the deeper (primary) eclipse in a light curve and $T_{0}$ is adopted as the moment of a deeper eclipse. Temperatures of the stars whose contribution to the total light of the system is greater (for ASAS-010538 it is the secondary, for ASAS-182510 - primary, for ASAS-182525 - primary) were obtained from the colour-temperature calibration by \citet{wor11}. We used colour values from the TYCHO-2 catalogue \citep{hog00}. These temperatures were fixed when calculating the temperatures of the other components of the systems using the \textsc{phoebe} code. The uncertainties of temperatures of the stars with a greater contribution to the total light of the system are based on the colour-temperature error propagation \citep{wor11}, the error budget in the case of the other components is based on both the uncertainties obtained in \textsc{phoebe} and the errors of colour-based temperatures. 

The photometric scale factors were adjusted in \textsc{phoebe}. Gravitational darkening coefficients were set at the value of $\beta=0.32$ \citep{luc67}, and a limb darkening effect was assumed to be based on the logarithmic law \citep{kli70} with the values of coefficients from the tables of \citet{van93}. We checked the presence of the third light fitting it in \textsc{phoebe} and as for ASAS-010538 and ASAS-182525 the result was statistically insignificant  (the third light values were less than its errors), we assumed there is no third light and applied it in the further analysis (the case of ASAS-182510 is discussed in Section 5.2). The surface albedo values were assumed to be the default values from \textsc{phoebe}. 

The photometric and spectroscopic solutions are presented in Figures 1--6 for the systems ASAS-010538, ASAS-182510 and ASAS-182525, respectively. Colour plots are available in the electronic version of the paper.

\begin{figure*}
	\begin{center}
	\begin{tabular}{cc}
	\includegraphics[scale=0.3, angle=-90]{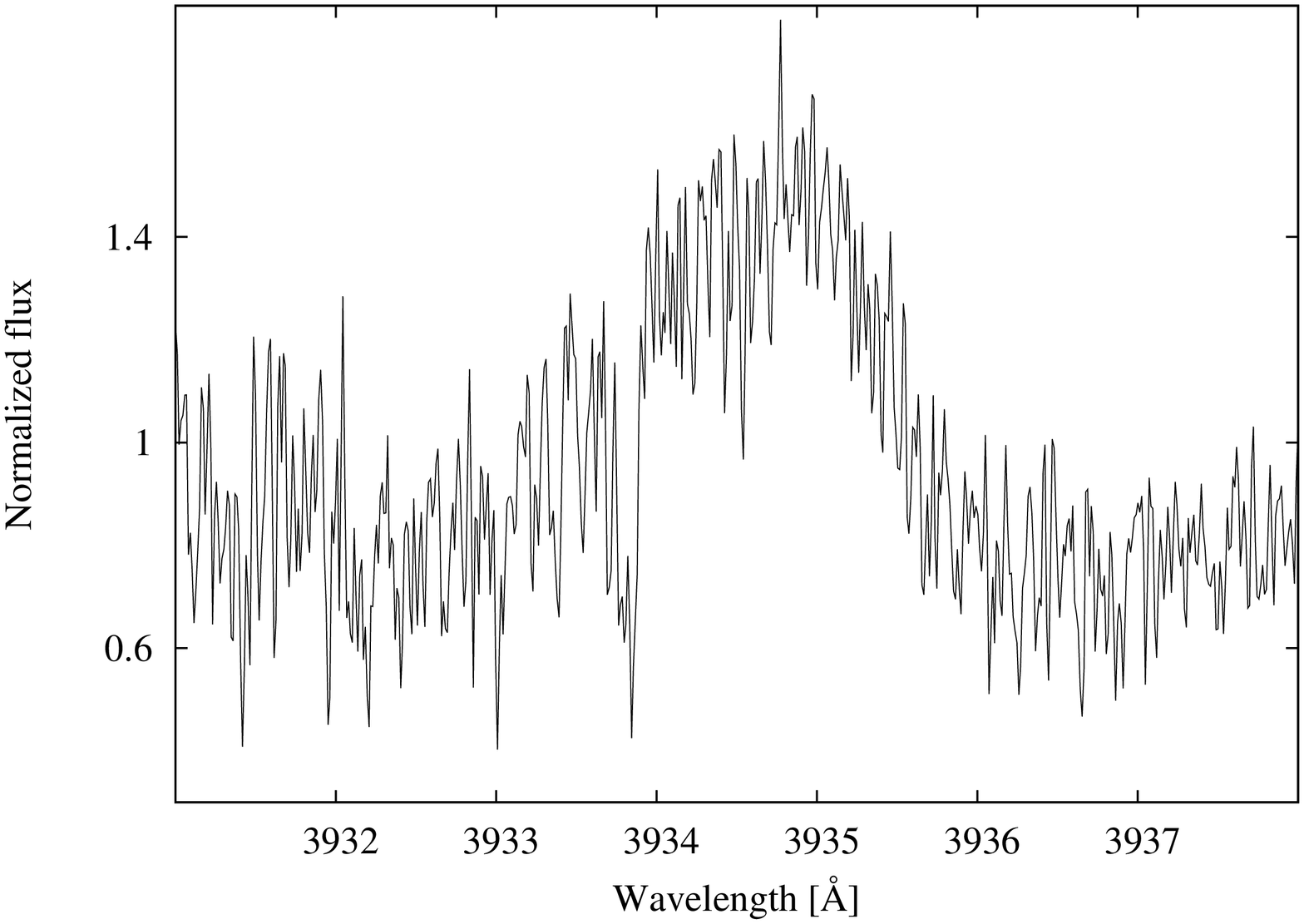}&
	\includegraphics[scale=0.3, angle=-90]{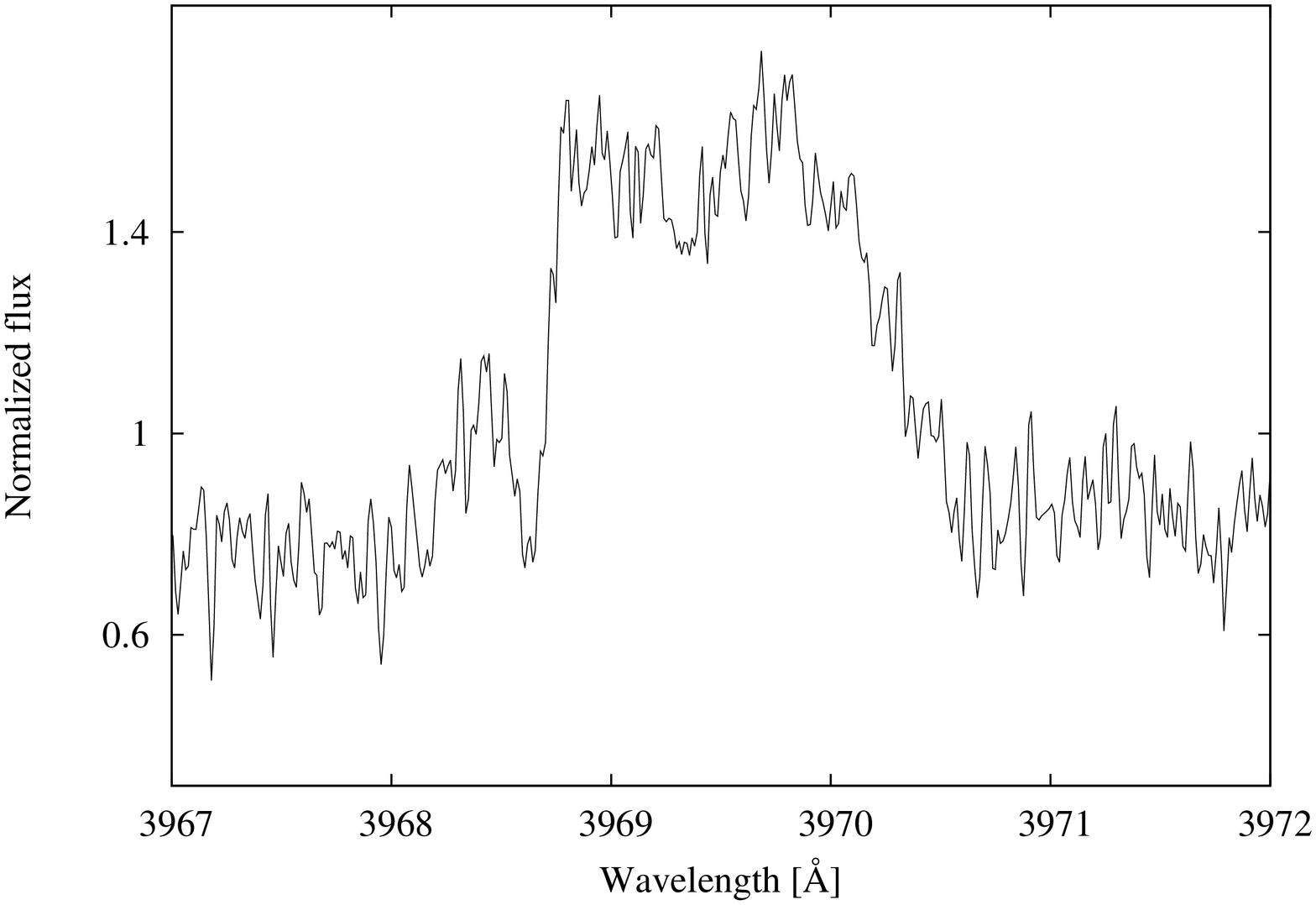}\\
	\end{tabular}
\caption{Spectral ranges of continuum-normalized spectra of ASAS-182525 in the core of Ca \textsc{II} K (left panel) and H (right panel) lines. Emission in these lines is clear and corresponds to the chromospheric activity of the components.}
\label{ca_emission}	
	\end{center}
\end{figure*}	

\subsection{ASAS-010358}
The resulting \textit{rms} in the radial velocities are 1.289 km/s and 3.488 km/s for the primary and secondary of ASAS-010538, respectively. The higher value of secondary's rms corresponds to the higher rotational velocity of the star which causes spectral line broadening.  The average photometric error is 0.034 mag, and the LC \textit{rms} is 0.033 mag.
 
The system's components are $\sim$1.4 more massive than the Sun, with radii of 2 R$_\odot$ and 5.3 R$_\odot$. A significant difference between our result and radii determined by \citet[][3.14 M$_\odot$ and 4.06 M$_\odot$]{hel09} is caused by more data points we used in our analysis and thus the ratio of the radii is better constrained. We would like to stress that for ASAS-010538 we derived the relative errors in the masses ($\Delta$M/M) and radii ($\Delta$R/R) at the 2$\%$, 3.6$\%$, and 3$\%$, 14$\%$ levels. The improved values of the period $P$, an ephemeris timebase $T_{0}$, alongside with the RV amplitudes $K_{1}$ and $K_{2}$, systemic velocity $\gamma$, inclination $i$, semimajor axis $a$, rotational velocities, and temperatures are presented in Table \ref{tab_orb}. Despite comparable masses we noticed a significant difference in the components' temperatures, which is a manifestation of the slightly different evolutionary stages of each star. 

\subsection{ASAS-182510}
The resulting RV \textit{rms}: 0.263 km/s and 0.363 km/s  for the primary and secondary, respectively, is significantly lower than for the ASAS-010358 system because of the smaller uncertainties of radial velocities due to lower values of rotational velocities of the components.  The average photometric error is 0.028 mag, and the LC \textit{rms} is 0.027 mag.

During the photometric analysis we realized that a close visual companion affected the ASAS light curve, so the third light contribution was taken into consideration. The ASAS apparent \textit{V}-band magnitude of 10.56 mag was assumed as the magnitude of the binary and the companion. We adopted the magnitude of the binary system \textit{V} = 10.87 mag from the Tycho-2 catalogue \citep{hog00} and calculated intensities of both the binary and the companion. The fraction of total light of the system due to the third body was determined to be $0.25$ and fixed in the analysis. 

We note that the components of ASAS-182510 are the most massive in the analyzed sample. The masses are almost equal and come to $\sim$ 3.3 M$_\odot$ while the radii are 23.4 R$_\odot$ and 15.3 R$_\odot$, respectively. The derived relative errors in the masses ($\Delta$M/M) and radii ($\Delta$R/R) are 1$\%$, 1$\%$, 13$\%$, and 24 $\%$, respectively. The results of our detailed analysis of the system are presented in Table \ref{tab_orb}.

\subsection{ASAS-182525}
We noticed out-of-eclipse and time-varying brightness modulations in the ASAS-182525 light curve. The detailed spectral studies yielded the detection of strong Ca \textsc{II} H (3\,969 \AA) and K (3\,934 \AA) emission lines which can indicate the presence of an active chromosphere for late-type stars \citep{str94}. The emission was detected in both spectral regions in the ASAS-182525 spectrum obtained by the HDS spectrograph (spectrum with the highest SNR for this star). The corresponding spectral regions are presented in Fig. \ref{ca_emission}. The analyzed spectrum was obtained at the orbital phase $\phi$=0.16 so emission features from both components are blended. We have also studied emission in the Balmer lines - although H$_\alpha$ and H$_\beta$ are outside of the HDS spectral region, we detected variability of the H$_\alpha$ emission lines in CHIRON and CORALIE spectra. These activity indicators imply the presence of spots on the stellar surfaces which cause brightness modulations. 

\begin{figure*}
	\begin{center}
	\begin{tabular}{cc}
	\includegraphics[scale=0.24, angle=-90]{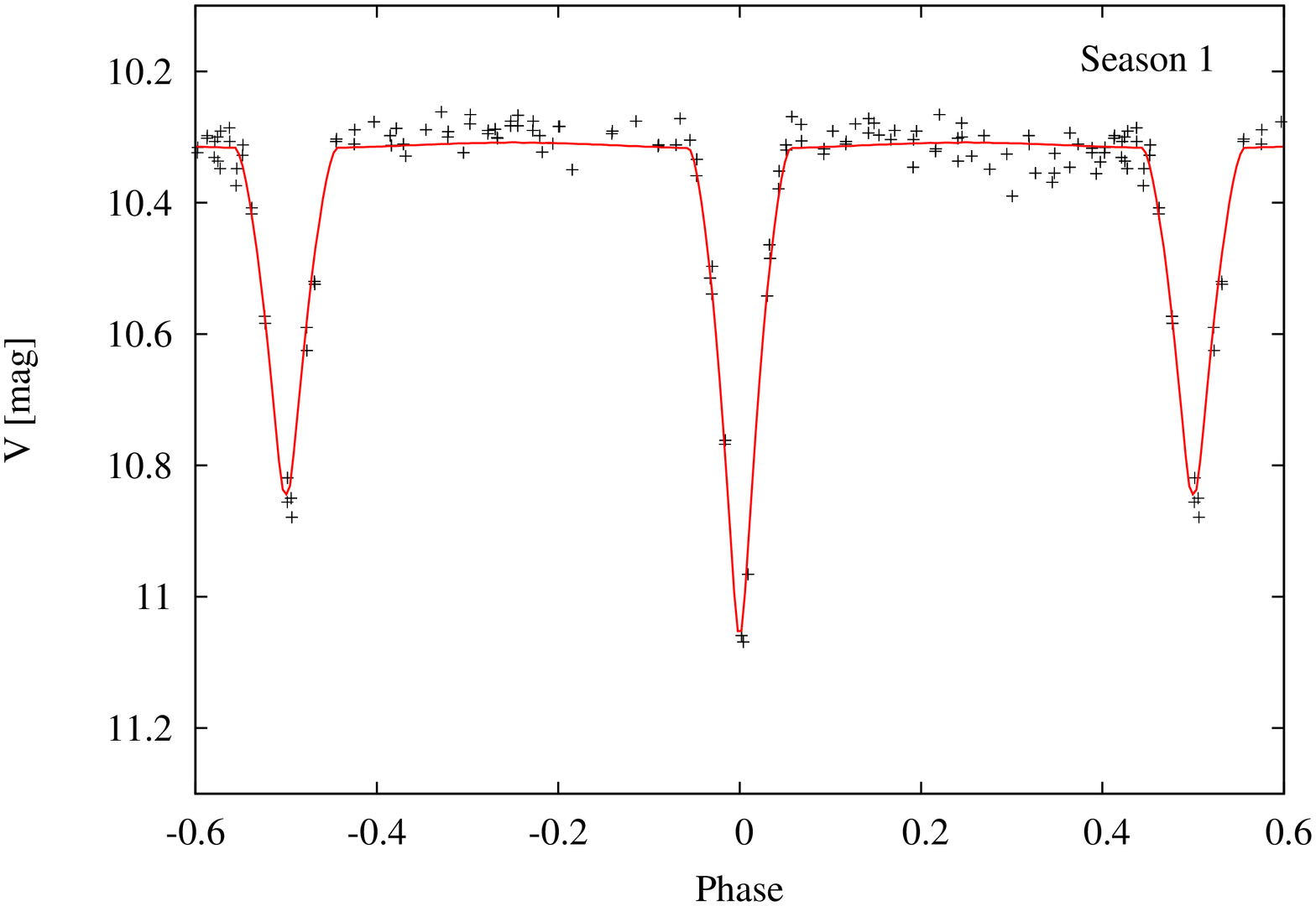}&
	\includegraphics[scale=0.24, angle=-90]{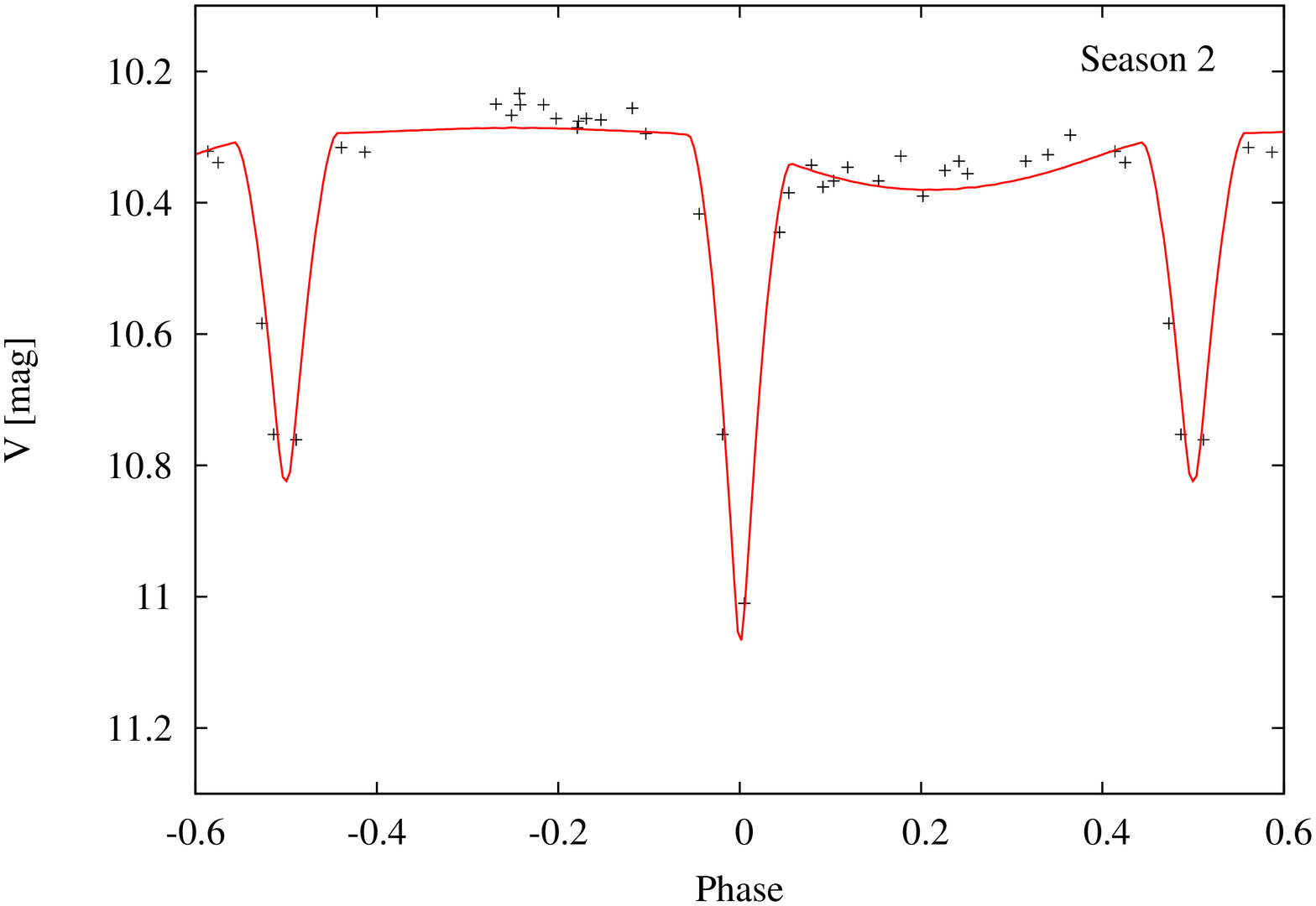}\\
	\includegraphics[scale=0.24, angle=-90]{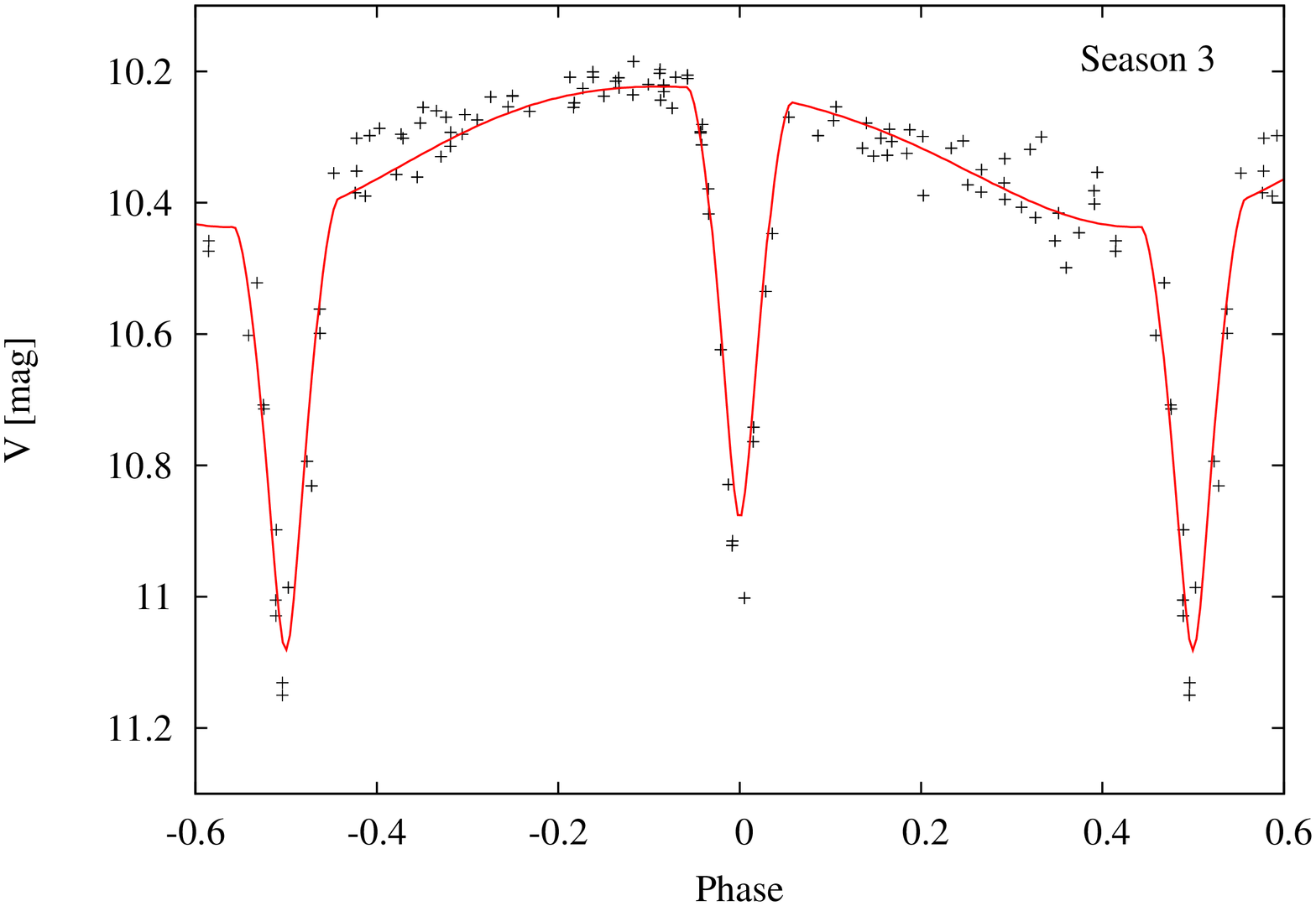}&
	\includegraphics[scale=0.24,angle=-90]{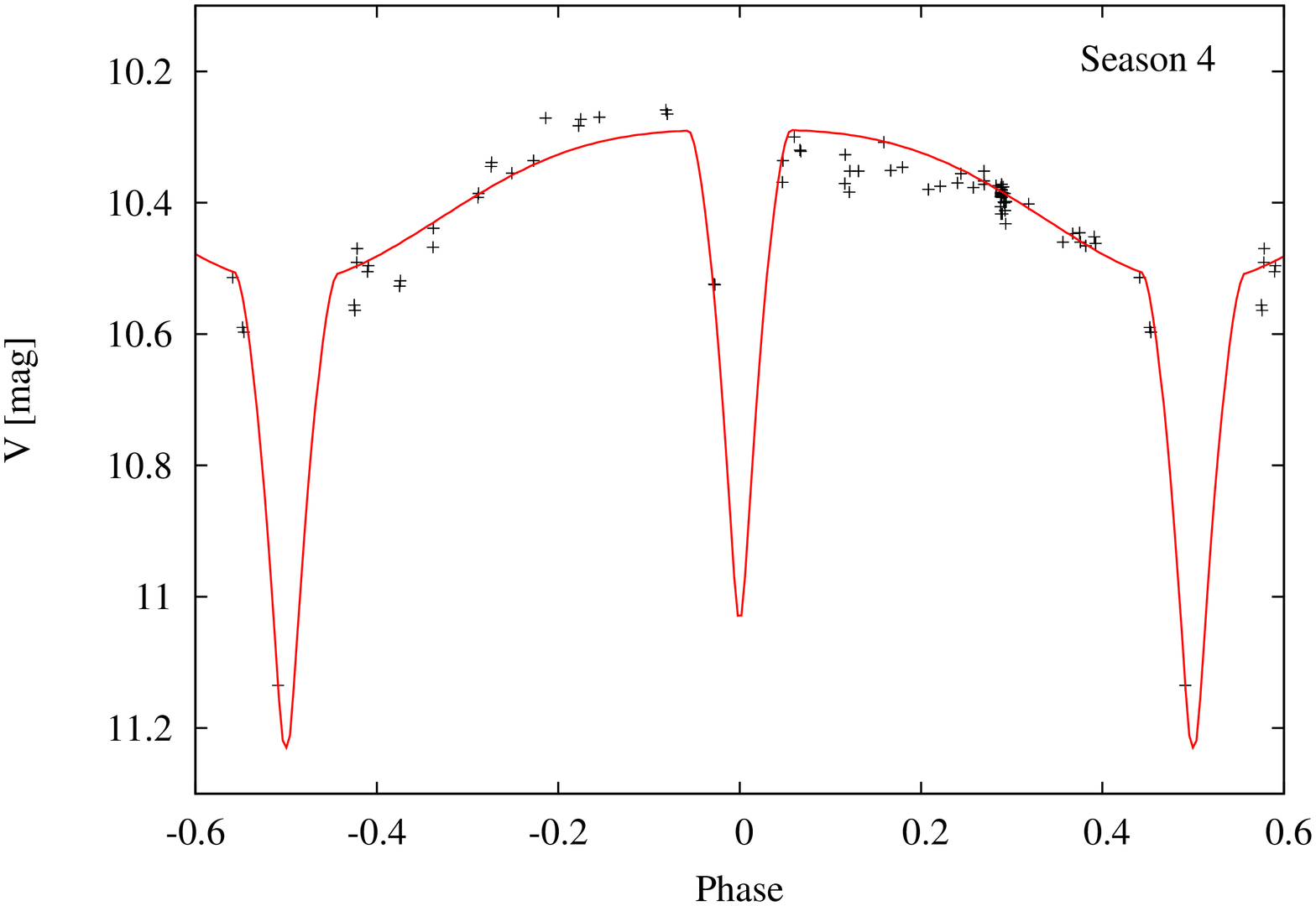}\\
	\includegraphics[scale=0.24,angle=-90]{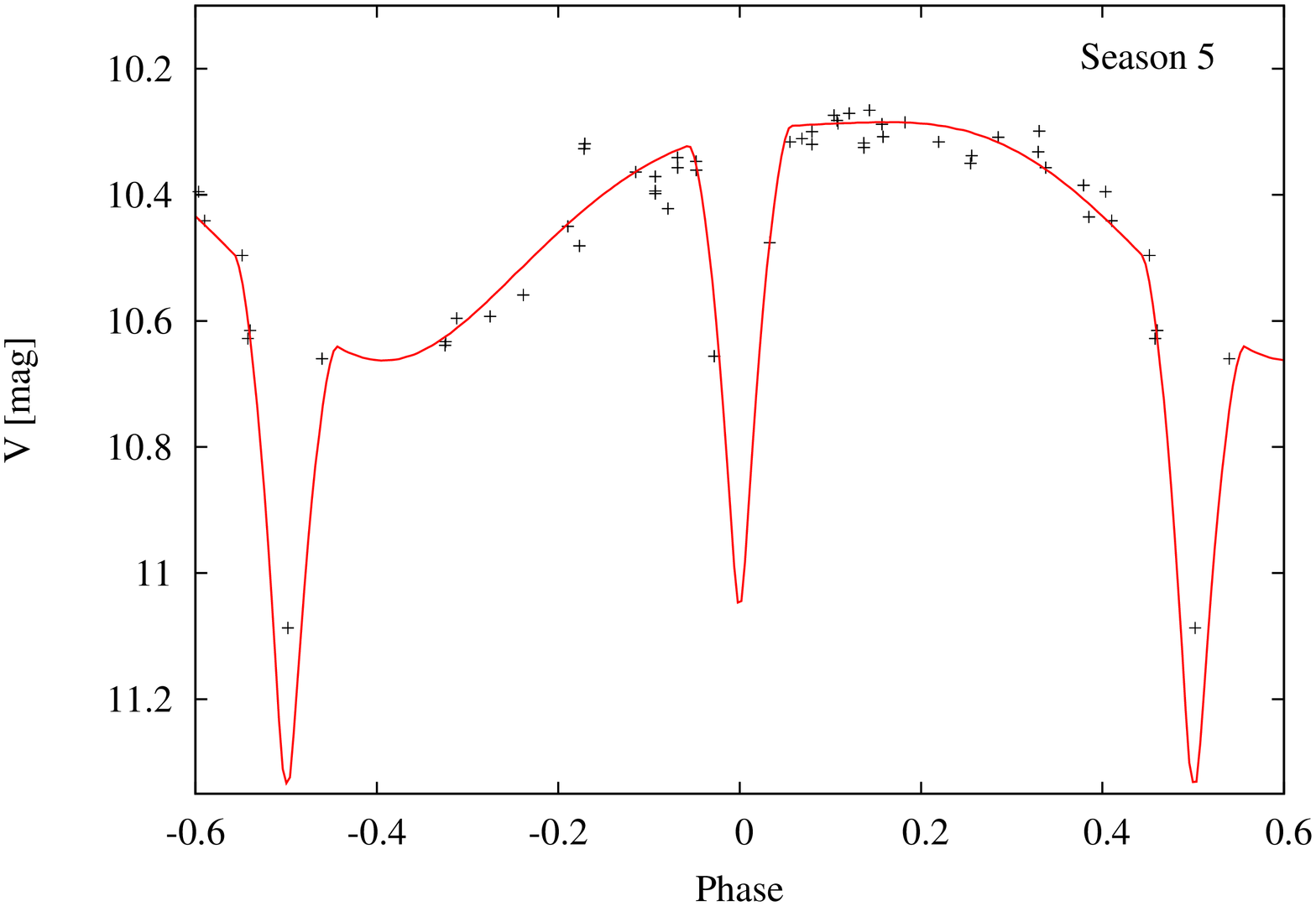}&
	\includegraphics[scale=0.24,angle=-90]{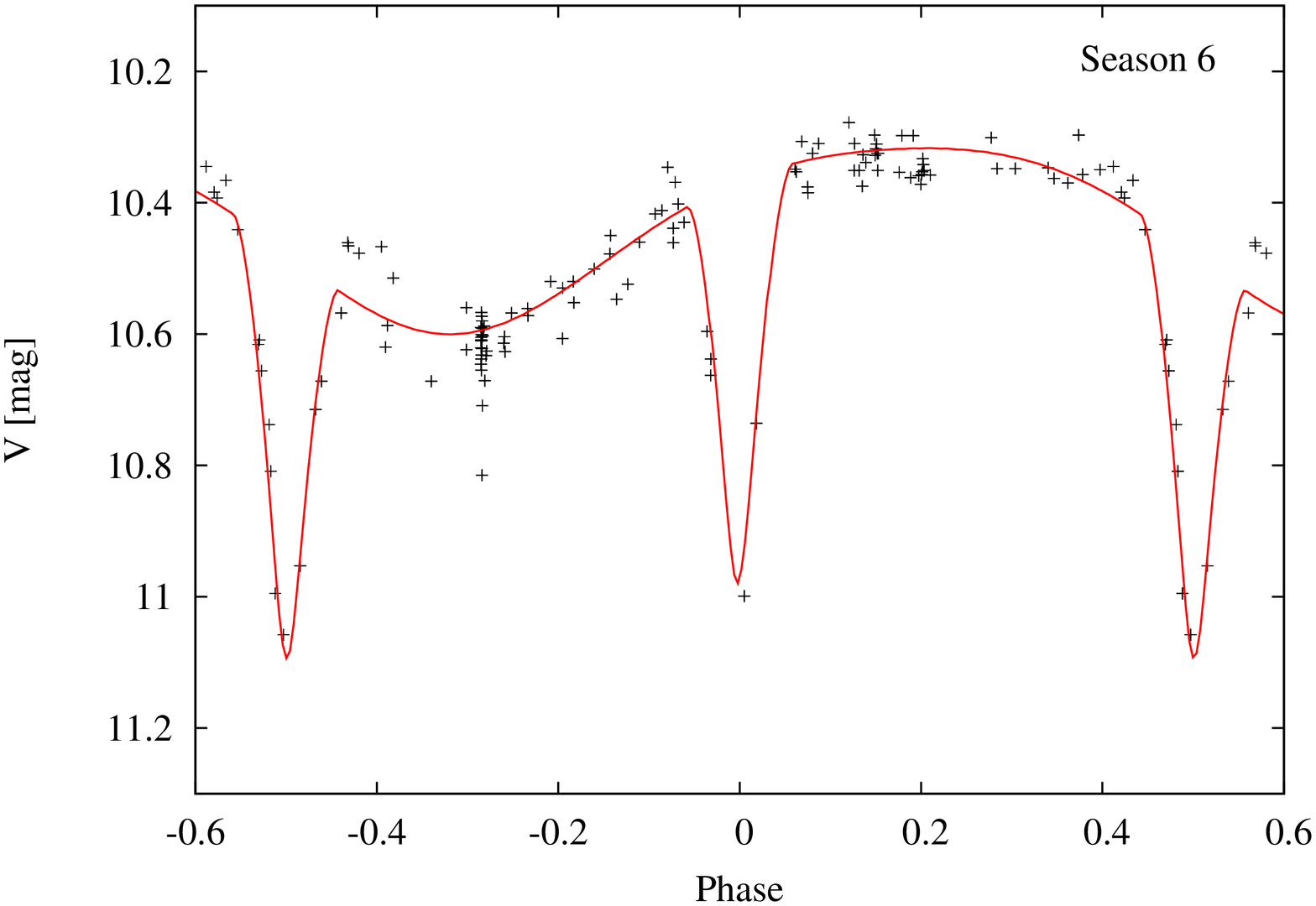}\\
	\includegraphics[scale=0.24,angle=-90]{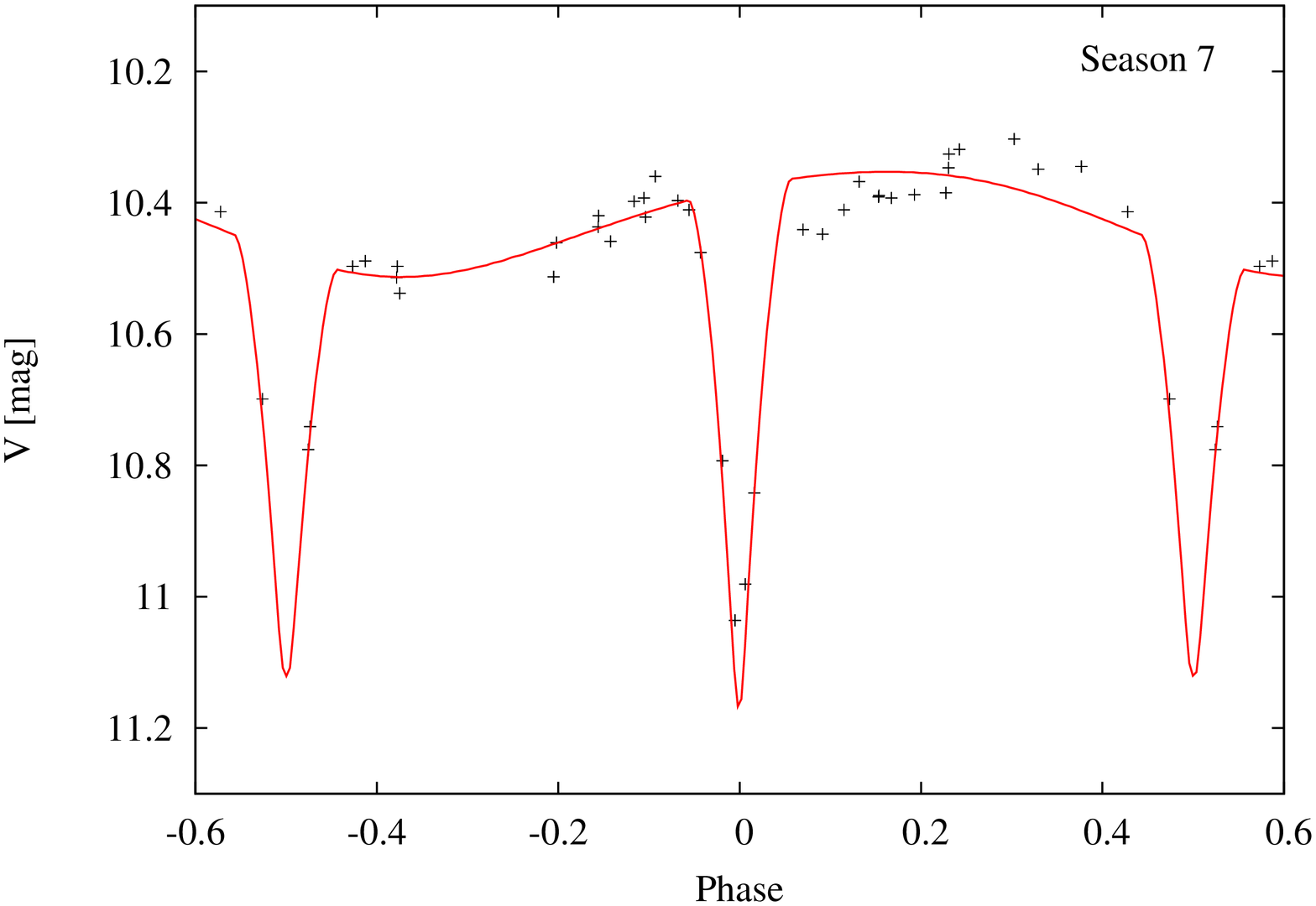}&
	\includegraphics[scale=0.24,angle=-90]{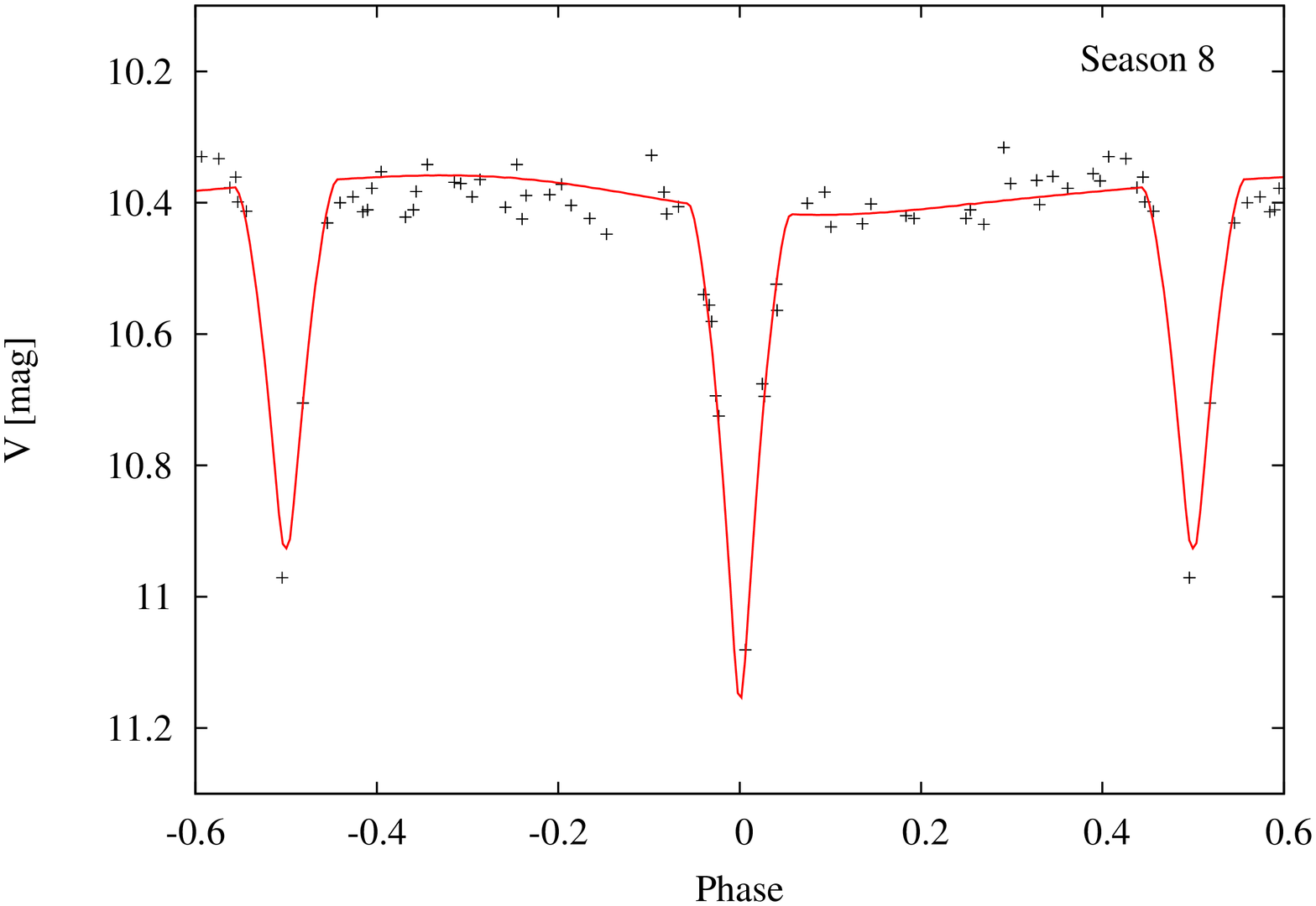}\\
	\includegraphics[scale=0.24,angle=-90]{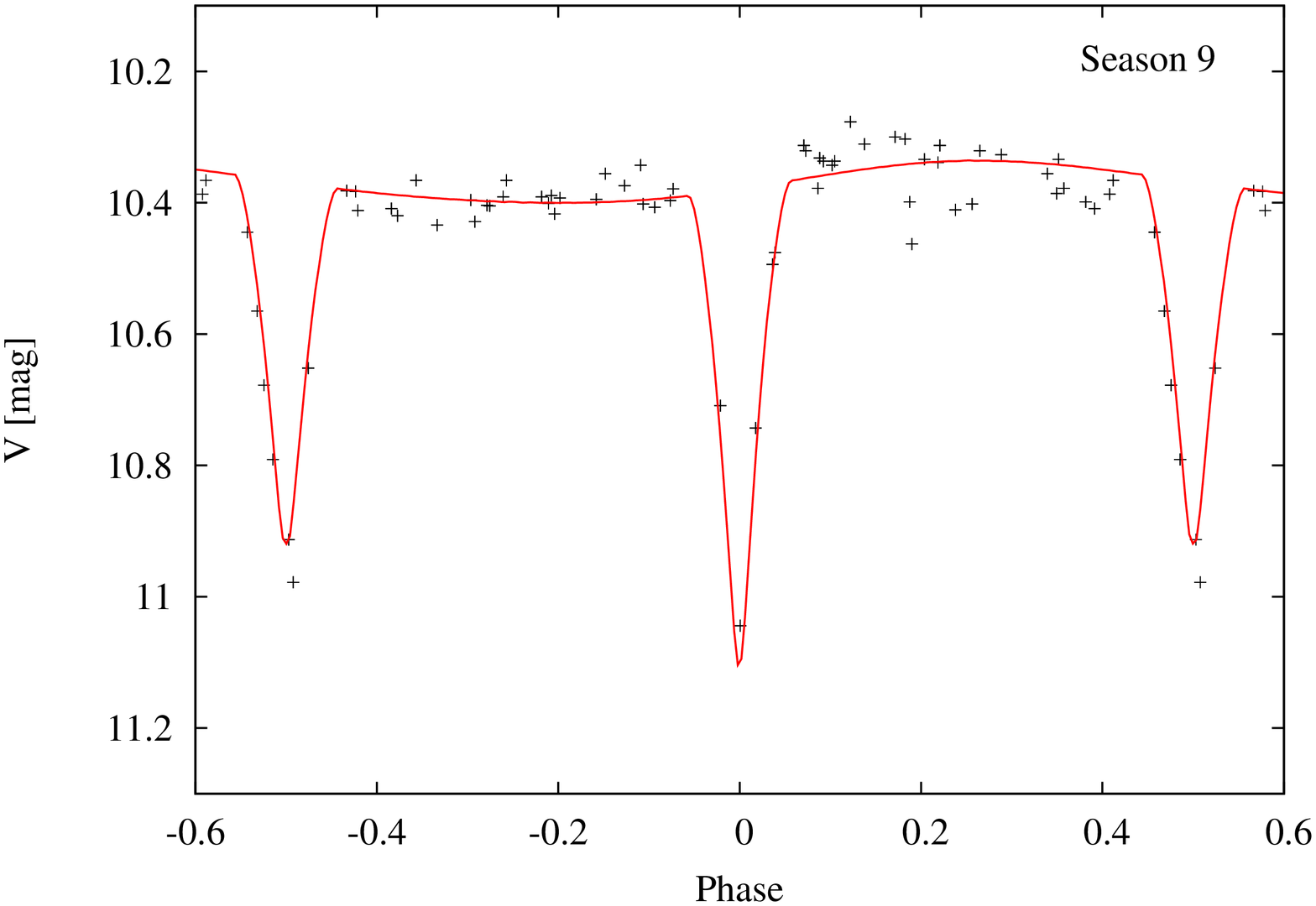}&
	\includegraphics[scale=0.24,angle=-90]{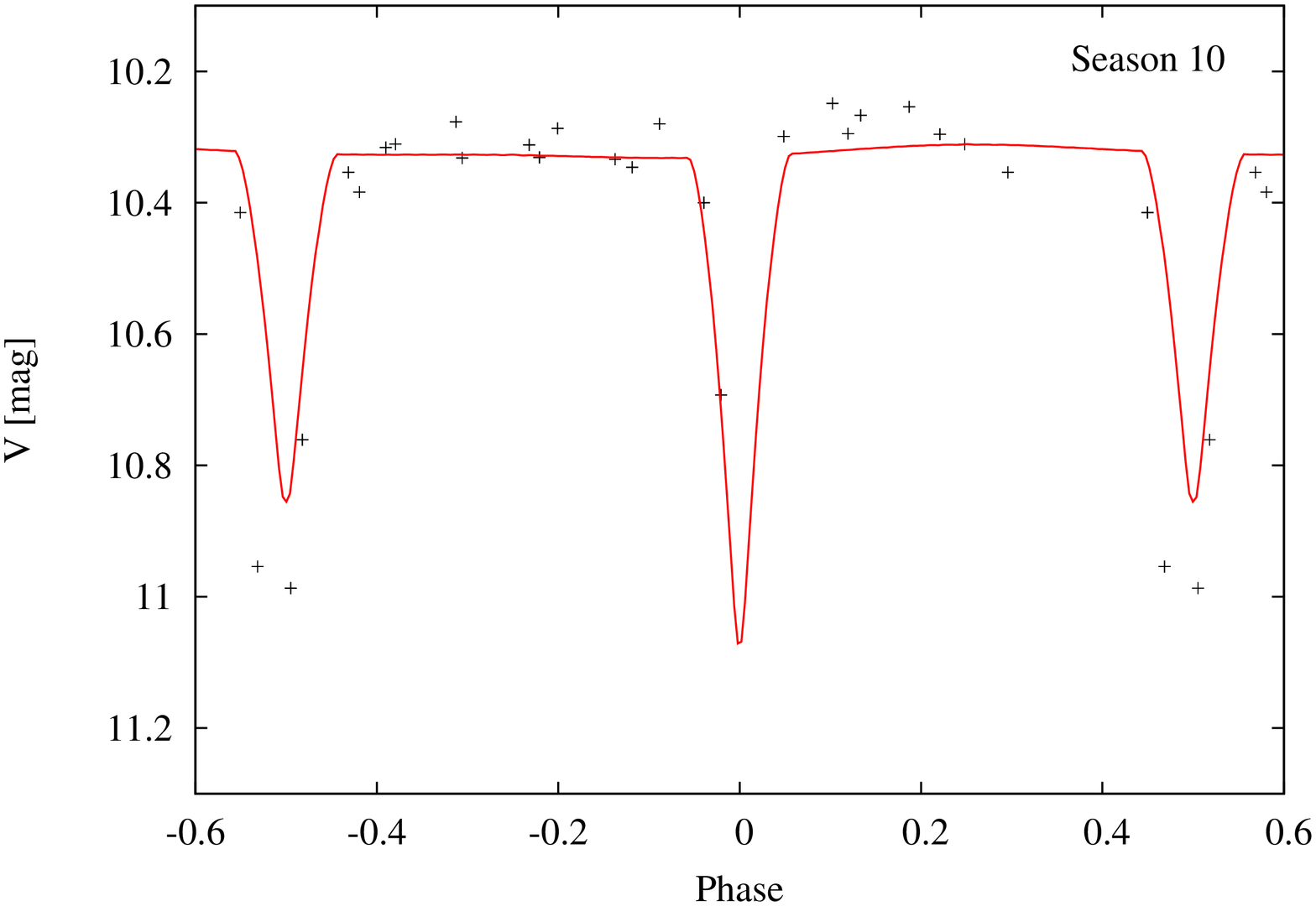}\\
	\end{tabular}
	
	\caption{Light curves of 10 seasons for ASAS-182525 with the best-fitting model. The influence of an evolving spot on the light curves is significant.} 
	\label{LC_182525_epochs}
	\end{center}
\end{figure*}

Splitting the data into 10 seasons allowed us to investigate the evolution of the spots. The analysis of the spots was carried out in \textsc{phoebe}. The approximated model assumes the existence of one spot on the surface of the system's primary component. The evidence for that is the fact that the depth of the secondary eclipse changes (when the primary star is closer to the observer). We observed that season 1 has no significant out-of-eclipse changes in brightness so we adopted it as a base season. We applied \textit{P} and $T_{0}$ derived from \textsc{jktebop} analysis of the full set of data to \textsc{phoebe} and estimated the physical and orbital parameters of the system based on the available spectroscopy and photometry just from season 1, assuming the starting value of inclination taken from \textsc{jktebop} analysis. The derived quantities were fixed in the analysis of the other seasons. We adjusted colatitude, longitude, radius and temperature of one spot located at the primary component's surface (Table \ref{spots_182525}). The adopted model with one cool spot reflects brightness changes in various seasons (see Fig. \ref{LC_182525_epochs}). Later we adjusted \textit{a}, $\gamma$, \textit{i}, temperatures and potentials in seasons 3, 4, 6, 9 (those seasons in which the brightness modulations are significant), calculated the weighted mean of those quantities and their errors and applied it in the further analysis. 

The derived orbital and physical parameters of ASAS-182525 are presented in Table \ref{tab_orb}.  The resulting \textit{rms} in radial velocities are 0.998 km/s  and 0.771 km/s for the primary and secondary of ASAS-182525, respectively. The average photometric error is 0.098 mag. 

The analysed system's components have almost equal masses of $\sim$ 1.2 M$_\odot$ and radii of 13 R$_\odot$ and 12.7 R$_\odot$.  The derived relative errors in the masses ($\Delta$M/M) and radii ($\Delta$R/R) are 3$\%$, 3$\%$, and 8$\%$, 8$\%$, respectively. The physical characteristics of the components and orbital parameters of the ASAS-182525 system are presented in Table \ref{tab_orb}.

We emphasise that the modulation in brightness of the whole system in seasons 5 and 6, where the influence of spots is significant, reaches $\sim$ 0.31 mag in comparison to season 1, which corresponds to a 25 $\%$ decrease in the total flux of the system. That means the spot would cause a $\sim$ 44 $\%$ decrease in the flux of the primary component of ASAS-182525. This is the first known case where the spots (here approximated by one spot) block almost half of the stellar flux.

\section{Discussion}
\subsection{Evolutionary status and age determination}

\begin{figure}
\includegraphics[scale=0.5,angle=270]{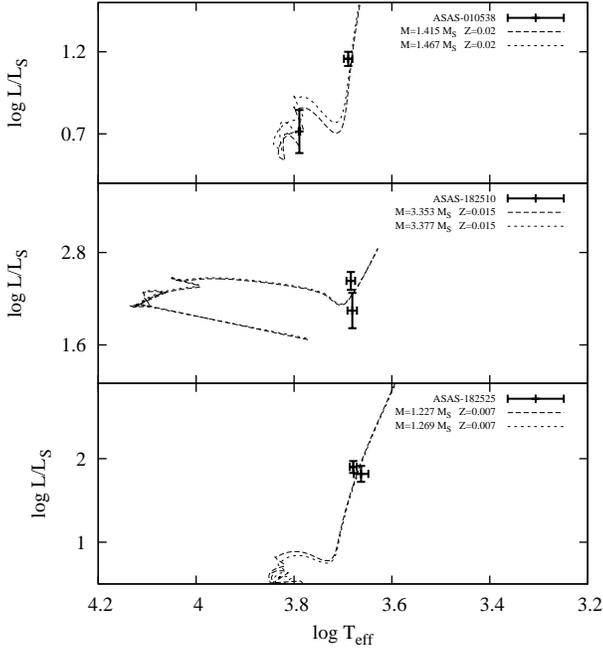}
\caption{YY evolutionary tracks for components of ASAS-010538, ASAS-182510, ASAS-182525} 
\label{track_kompil}
\end{figure}

To check the evolutionary status of the selected systems we compared our results with the Yonsei-Yale stellar evolution models \citep[hereafter YY,][]{yi01}.  Evolutionary tracks interpolated for the determined masses and $\alpha$-enhancement of zero are presented in Fig. \ref{track_kompil}. As the tracks for solar metallicity have matched only observations for ASAS-010538 system, we adopted the metallicity of Z=0.015 for ASAS-182510, and Z=0.007 for ASAS-182525. 

We can clearly estimate the status of all of the analyzed systems - stars belonging to ASAS-182510 and ASAS-182525 are located at the red giant branch. ASAS-010538 is a system with components at slightly different phases of evolution - the secondary is a red giant while the primary has evolved away from the main sequence and appears to be at a subgiant phase.  

\begin{figure}
\includegraphics[scale=0.5,angle=270]{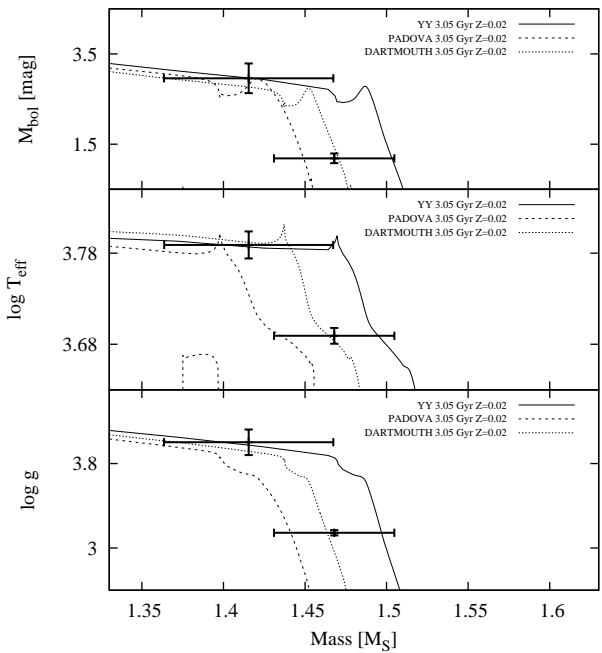}
\caption{YY, Padova, and Dartmouth isochrones for 010538} 
\label{010538_iso_kompil}
\end{figure}

\begin{figure}
\includegraphics[scale=0.5,angle=270]{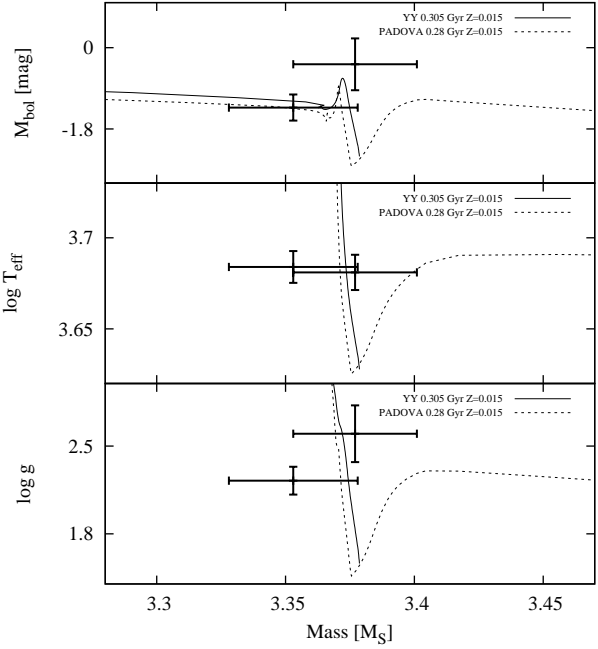}
\caption{YY and Padova isochrones for 182510} 
\label{182510_iso_kompil}
\end{figure}

\begin{figure}
\includegraphics[scale=0.5,angle=270]{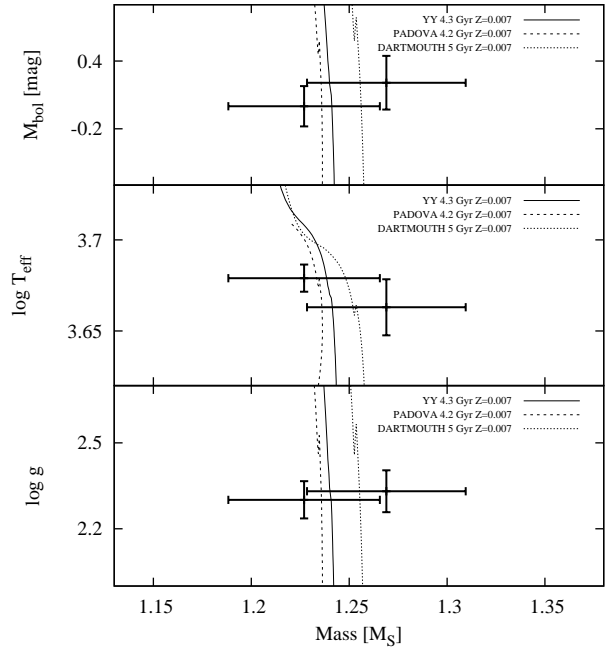}
\caption{YY, Padova, and Dartmouth isochrones for 182525} 
\label{182525_iso_kompil}
\end{figure}

To estimate the age of the systems we compared the results with three sets of widely used isochrones: YY \citep{yi01,dem04}, Padova \citep{gir00, mar08}, and Dartmouth \citep{dot07} assuming the metallicity we obtained from evolutionary status analysis (Z=0.02 for ASAS-010538, Z=0.015 for ASAS-182510, and Z=0.007 for ASAS-182525) and $\alpha$-enhancement of zero. We determined the location of the systems' components in three planes: M$_{bol}$ - Mass, log T$_{eff}$ - Mass, and log g - Mass and checked if a single isochrone fits both stars of the system simultaneously. One can find the plots for every system in Fig. \ref{010538_iso_kompil} - \ref{182525_iso_kompil}.

We found the age of ASAS-010538 to be around 3.05 Gyr (see Fig. \ref{010538_iso_kompil}). Isochrones of that age generated by all YY, Padova and Dartmouth codes are consistent with the observational data.

ASAS-182510 results were compared with just two sets of isochrones (YY and Padova; Dartmouth codes start with the age of 1 Gyr). YY gives a consistent for both components result of 0.305 Gyr but we encounter problems reproducing such a value using the Padova sets of isochrones which indicate the age of the system to be around 0.28 Gyr (see Fig. \ref{182510_iso_kompil}). The discrepancy between age estimation originates in the different definition of the zero-age (the ages computed in YY models include also a pre-MS stage) and differences in physics adopted for certain models. 

We estimated the age of ASAS-182525 to be around 4.3 Gyr (YY), 4.2 Gyr (Padova), 5 Gyr (Dartmouth; see Fig. \ref{182525_iso_kompil}). The reason of the disagreement between values obtained using various codes is the same as for ASAS-182510.

It is worth mentioning that the temperature of systems' components which have the greater contribution to the total light of the system were obtained from the colour-temperature calibration and are based on the colours of the whole system, what is an additional source of uncertainty in temperatures determination. In order to avoid this issue multi-colour photometry is essential. 

\begin{table}
\caption{Parameters of a spot on the primary component's surface of ASAS-182525 in various seasons.}
\label{spots_182525}
\centering
\begin{tabular}{l c c c c }
\hline \hline 
\scriptsize {Season} & \scriptsize {Colat. [deg]} & \scriptsize {Long. [deg}] & \scriptsize{Radius [deg]} & \scriptsize{Temp. [$T$/$T_\odot$]}\\
\hline
1 & -- & -- & -- &  --\\
2 & 75 & 281 & 23 & 0.85\\
3 & 22 & 204 & 66 & 0.85\\
4 & 40 & 179 & 58 & 0.89\\
5 & 47 & 140 & 56 & 0.79\\
6 & 37 & 114 & 56 & 0.75\\
7 & 19 & 130 & 56 & 0.79\\
8 & 10 & 314 & 56 & 0.90\\
9 & 29 & 81 & 75 & 0.97\\
10 & 18 & 73 & 45 & 0.97\\
\hline
\end{tabular}
\end{table}

\subsection{Distance determination}

One can derive the distances to the analyzed systems using various methods. The traditional method uses bolometric corrections to find the absolute visual magnitude. The luminosity of each star was calculated from its radii and effective temperature, and the obtained values of absolute bolometric magnitudes were converted to absolute visual magnitudes using the bolometric corrections of \citet{bes98}. The combined absolute visual magnitude of the two stars was then compared to the apparent visual magnitude in order to find the distance to the system. The estimation of the reddening was performed using maps of dust infrared emission by \citet{sch98}. We obtained colour excesses $E(B-V)$ of 0.061 mag, 0.44 mag, and 0.45 mag for ASAS-010538, ASAS-182510, and ASAS-182525, respectively. The distance determination procedure was performed in \textsc{jktabsdim} and the results are as follows: $d_{A010538}$ = 459 $\pm$  15 pc, $d_{A182510}$ = 1758 $\pm$ 232 pc, $d_{A182525}$ =  765 $\pm$ 45 pc. 

The second method \citep{sou05} allows us to determine the distance to ED system from the use of empirical relations between surface brightness and effective temperatures, derived from interferometry by \cite{ker04}. However, the established laws between angular sizes and effective temperatures are valid only for main sequence stars and subgiants. We tried to use that method for ASAS-010538 which consists of a subgiant and a giant but the derived distance was not consistent with the value obtained from the use of bolometric corrections.

An alternative method which can be used in the case of giant stars uses the relations between surface brightness and colour indices. That solution was successfully used by \citet{pie09} and \citet{gra12} to determine distances to DEBs composed of evolved giants in Magellanic Clouds. They used the calibration of \citet{diB05} between \textit{V}-band surface and \textit{V}-\textit{K} colour relation obtained from a sample of giant and dwarf stars for which the angular diameters were measured using interferometry. However because we have no multi-colour photometry for any of the analyzed systems, we could not employ such a method for distance determination.   

Parallaxes and hence distances to any of the analyzed systems have not been derived before so we were not able to compare our results with the literature.

\section{Conclusions}

We present for the first time RV curves derived from high-resolution spectra, for both components of the systems ASAS-182510 and ASAS-182525. These, accompanied by a detailed analysis of the ASAS \textit{V}-band photometry, have allowed us to obtain accurate orbit solutions and fundamental physical parameters of the components. We have also studied a previously analyzed system ASAS-010538 for which we obtained an improved orbital solution and more accurate parameters. 

The masses of each of the systems' components were determined with an accuracy of $\sim$ 3 $\%$ or better, but we have not reached the required 3 $\%$ precision in radii.  However, the results are still highly valuable for performing a reliable estimation of the evolutionary status of the components. We compared the observations with stellar evolution models. We find that five stars are within the error bars located on the red giants branch and one is a subgiant. The estimated ages are about 3 Gyr, 0.3 Gyr, and 4.5 Gyr for ASAS-010538, ASAS-182510, and ASAS-182525, respectively. We have not found significant differences between the empirical data and theoretical models. We have fitted single isochrones that match both components of the analyzed systems. 

In order to improve the results and reach the required 3$\%$ precision in radii, precise photometric measurements for all of the systems are required. With the addition of multicolour photometry one would be able to determine radii and temperatures with a much higher precision. Moreover, the assumed metallicity of the systems is based on an analysis involving theoretical evolutionary tracks. With more available spectra we could perform a spectral disentangling and analyze individual spectra of all of the components separately to determine stellar chemical abundances. The analyzed systems are definitely interesting objects for further studies mostly because of their evolutionary status, but also because of activity (chromospheric activity indicators were found for ASAS-182525 system). Additional photometric and spectroscopic data would allow the analyzed stars to become demanding test beds of stellar evolution theories. 

\section*{Acknowledgments}

The authors would like to thank H. Worters for support during our observations at SAAO, A. Mazierska for help in RV calculations, A. Smith and V. Ribeiro for comments and suggestions.

This work is supported by the National Science Center through grants 2011/01/N/ST9/02209 (MR), 5813/B/H03/2011/40 (MK), and 2011/03/N/ST9/01819 (KGH). KGH acknowledges support provided by the Proyecto FONDECYT Postdoctoral No. 3120153. MK is supported by the European Research Council through a Starting Grant and the Foundation for Polish Science through "Idee dla Polski" funding scheme. AJ acknowledges support from Fondecyt project  1130857 and the Millennium Science Initiative, Chilean Ministry of Economy (Nucleus P10-022-F).

This paper uses observations made at the South African Astronomical Observatory (SAAO). This research has made use of the Simbad database, operated at CDS, Strasbourg, France. 

The authors wish to recognize and acknowledge the very significant cultural role and reverence that the summit of Mauna Kea has always had within the indigenous Hawaiian community. We are most fortunate to have the opportunity to conduct observations from this mountain.

\appendix

\section{RV measurements for ASAS-010538, ASAS-182510, ASAS-182525 systems}

The section includes Tables \ref{RV_010538}--\ref{RV_182525} with RV measurements, formal RV errors, O-Cs, exposure times for each spectrum, SNR and telescope specifications for both components of the selected systems. The used telescopes/spectrographs are as follows: R/G = Radcliffe/GIRAFFE, ESO/H = ESO 3.6~m/HARPS, CTIO/CH = CTIO 1.5~m/CHIRON, EUL/C = Euler/CORALIE, SUB/HDS = Subaru/HDS (red or blue CCD chip). SNR stands for a signal-to-noise ratio per collapsed spectral pixel at $\lambda$=5\,500 \AA.

\begin{table*}
\caption{RV measurements for ASAS-010538.}
\centering
\begin{tabular}{c c c c c c c c c c}
\hline \hline
BJD-2450000 & $RV_{1}$ & $\sigma_{RV_{1}}$ & $O-C_{1}$ & $RV_{2}$ & $\sigma_{RV_{2}}$ & $O-C_{2}$ & $T_{exp}$ & SNR & Tel./Sp.\\
 & [km/s] & [km/s] & [km/s] & [km/s] & [km/s] & [km/s] & [s]\\
\hline
 4007.418443 &-25.896 &1.170  & 1.590 & 13.767 & 2.499 & 2.095 & 3600 & 18 & R/G\\
 4008.569095 & 34.786 & 2.262 &-2.264 & -48.402 & 2.337 &  2.147  & 3600 & 28 & R/G\\
 4166.269636 & -63.431 & 6.604 & 3.312 & 54.048 & 6.154 & 4.502 & 3600 & 10 & R/G\\
 5435.568405 & -36.676 & 2.215 & -0.011 & 19.068 & 7.345  &-1.458 & 2700 & 40 & R/G \\
 5440.547112 & -32.701 &	0.803  &  0.136 & 17.075 & 8.869 & 0.242  & 3600 & 40 & R/G\\
 5469.588733 & 53.808	&3.437 & -0.776 & -59.849 & 3.609 &7.595 & 2700 & 10 & R/G\\
 5721.857804 &  30.650	 & 0.317  & -0.424 & -49.181 & 2.808   & -4.392 & 1200 & 50 & ESO/H\\
 5722.917250 & -28.711 & 0.299 & 0.484 & 15.516 & 1.583 & 2.195 & 1200 & 65 & ESO/H\\
 5811.700692 & -30.106 & 0.335  & 0.111 & 14.861 & 1.583 & 0.555 & 1200 & 35 & ESO/H\\
 5812.690316 & -74.520 & 0.294 &  -0.381 & 54.189 & 6.552  & -2.494 & 1500 & 45 & ESO/H\\
 5813.758491 & -79.118 & 0.281 & -0.063 & 61.284 & 1.630 & -0.145 & 1500 & 55 & ESO/H\\
 5815.875346 & 22.590 & 1.492  & -0.247 &-35.837 & 6.924 & 1.013 & 600 & 10 & CTIO/CH\\
 5823.816207 & 15.950 & 0.650 & 0.208 & -35.031 & 5.613 & -5.020 & 660 & 15 & CTIO/CH\\

\hline
\label{RV_010538}
\end{tabular}
\end{table*}

\begin{table*}
\caption{RV measurements for ASAS-182510.}
\centering
\begin{tabular}{c c c c c c c c c c}
\hline \hline
BJD-2450000 & $RV_{1}$ & $\sigma_{RV_{1}}$ & $O-C_{1}$ & $RV_{2}$ & $\sigma_{RV_{2}}$ & $O-C_{2}$ & $T_{exp}$ & SNR & Tel./Sp.\\
 & [km/s] & [km/s] & [km/s] & [km/s] & [km/s] & [km/s] & [s]\\
\hline

 5776.737886  & 48.344 & 0.480  &-0.699 & -21.434 & 0.243 &-0.304 & 720 & 10 & CTIO/CH\\
 5819.556752  & -21.876 & 0.230  & 0.024 & 50.210 & 0.205 & -0.113  & 720 & 15 & CTIO/CH\\
 5823.571475 & -12.426 & 0.169  & 0.146 & 40.717 & 0.287 &-0.207 & 720 & 15 & CTIO/CH\\
 5853.487853  & 58.619 & 0.674  &-0.565 & -31.873 & 0.331 &-0.535 & 720 & 20 & CTIO/CH\\
 5846.517685  & 52.704 & 0.214  &-0.112 & -24.896 & 0.191 & 0.032 & 600 & 20 & EUL/C\\
 6081.622312  & -17.273 & 0.213  & 0.030 & 45.540 & 0.226 &-0.151 & 600 & 20 & EUL/C\\
 5778.772055  & 44.664 & 0.149  & 0.134 & -16.570 & 0.144 & 0.016 & 600 & 70 & SUB/HDS red\\
 5778.772055  & 44.783 & 0.177 & 0.253& -16.421 & 0.239 & 0.165 & 600 & 60 & SUB/HDS blue\\
 
\hline
\label{RV_182510}
\end{tabular}
\end{table*}

\begin{table*}
\caption{RV measurements for ASAS-182525.}
\centering
\begin{tabular}{c c c c c c c c c c}
\hline \hline
BJD-2450000 & $RV_{1}$ & $\sigma_{RV_{1}}$ & $O-C_{1}$ & $RV_{2}$ & $\sigma_{RV_{2}}$ & $O-C_{2}$ & $T_{exp}$ & SNR & Tel./Sp.\\
 & [km/s] & [km/s] & [km/s] & [km/s] & [km/s] & [km/s] & [s]\\
\hline

 5776.726278 & 80.479 & 1.049 & 0.102 & 27.469 & 1.044 & 0.144 &  660 & 15 & CTIO/CH\\
 5815.550671 & 69.652 & 0.998 & -1.386 & 36.774 & 1.031 & 0.418 & 660 & 15 & CTIO/CH\\
 5823.592317 & 95.415 & 1.074 & -0.297 & 13.083  & 1.049 & 0.584 & 670 & 17 & CTIO/CH\\
 5846.531984 & 16.244 & 1.090 & -0.299  & 87.947  & 1.012 & -1.125 & 480 & 20 & EUL/C\\
 6080.665495 & 26.755 & 1.034 & 1.410 &81.073  & 1.019 & 0.518 & 480 & 15 & EUL/C\\
 5778.783498 & 89.888 & 1.007 & 0.681 & 18.039  & 0.990 & -0.749 & 600 & 70 & SUB/HDS red\\
 5778.783498 & 90.399 & 1.079 & 1.191 & 17.818  & 1.039 & -0.970 & 600 & 60 & SUB/HDS blue\\

\hline
\label{RV_182525}
\end{tabular}
\end{table*}

\bsp

\label{lastpage}

\end{document}